\documentclass[prb,twocolumn,groupedaddress,showpacs]{revtex4}

\usepackage{graphicx}
\usepackage{dcolumn}
\usepackage{amssymb}
\usepackage{bm}
\usepackage{epsfig}

\begin{document}

\title{Josephson junction detector of non-Gaussian noise}
\author{Hermann Grabert}

\affiliation{Physikalisches Institut and Freiburg Institute of
Advanced Studies, Albert-Ludwigs-Universit\"at, 79104 Freiburg,
Germany}

\date{February 22, 2008}

\begin{abstract}
The measurement of higher order cumulants of the current noise
generated by a nonlinear mesoscopic conductor using a Josephson
junction as on-chip detector is investigated theoretically.  The
paper addresses the regime where the noise of the mesoscopic
conductor initiates activated escape of the Josephson detector out
of the zero-voltage state, which can be observed as a voltage
rise.  It is shown that the deviations from Johnson-Nyquist noise
can mostly be accounted for by an effective temperature which
depends on the second noise cumulant of the conductor. The
deviations from Gaussian statistics lead to rather weak effects
and essentially only the third cumulant can be measured exploiting
the dependence of the corrections to the rate of escape from the
zero-voltage state on the direction of the bias current. These
corrections vanish as the bias current approaches the critical
current. The theory is based on a description of irreversible
processes and fluctuations in terms of state variables and
conjugate forces. This approach, going back to work by Onsager and
Machlup, is extended to account for non-Gaussian noise, and it is
shown that the thermodynamically conjugate force to the electric
charge plays a role similar to the counting field introduced in
more recent approaches to describe non-Gaussian noise statistics.
The theory allows to obtain asymptotically exact results for the
rate of escape in the weak noise limit for all values of the
damping strength of the Josephson detector. Also the feedback of
the detector on the noise generating conductor is fully taken into
account by treating both coupled mesoscopic devices on an equal
footing.
\end{abstract}


\maketitle
%%%%%%%%%%%%%%%%%%%%%%%%%%%%%%%%%%%%%%%%%%%%%%%%%
%%%%%%%%%%%%%%%%%%%%%%%%%%%%%%%%%%%%%%%%%%%%%%%%%
%%%%%%%%%%%%%%%%%%%%%%%%%%%%%%%%%%%%%%%%%%%%%%%%%
%%%%%%      I N T R O D U C T I O N       %%%%%%%
%%%%%%%%%%%%%%%%%%%%%%%%%%%%%%%%%%%%%%%%%%%%%%%%%
%%%%%%%%%%%%%%%%%%%%%%%%%%%%%%%%%%%%%%%%%%%%%%%%%
%%%%%%%%%%%%%%%%%%%%%%%%%%%%%%%%%%%%%%%%%%%%%%%%%
\section{Introduction}
Traditional nonequilibrium thermodynamics assumes Gaussian
fluctuations of the gross variables about their mean
values.\cite{LandauV} This assumption is a natural consequence of
the central limit theorem implying small fluctuations of additive
variables distributed in a Gaussian way. In the last decade there
have been extensive theoretical efforts\cite{Levitov,reviews} to
calculate deviations from Gaussian statistics for electronic
current fluctuations of mesoscopic devices. The complete knowledge
of the number of charges transferred through the device in a given
interval of time is referred to as full counting statistics (FCS).
It has turned out that FCS reveals details on microscopic
processes in the device that are not available through mere
measurements of the mean current and the noise variance. This can
already be seen from a simple example known since a long
time.\cite{Schottky} The FCS of a tunnel junction is Poissonian
when the applied voltage is large compared to the temperature ($eV
\gg k_BT$). In this case charges essentially only tunnel from
source to drain, and the Poissonian statistics points to
statistically independent transfers of discrete charges.

In contrast to the substantial literature on theoretical
predictions for FCS there are only rather few
experiments\cite{Reulet, Fujisaka, Reznikov, Ensslin, Pekola,
Pothier} that have measured deviations from Gaussian noise. This
is a consequence of the fact that these deviations are typically
small and require sophisticated experimental techniques to be
detected. The pioneering work by Reulet {\it et al.}\cite{Reulet},
has measured the third cumulant of the noise produced by a tunnel
junction. Since the noise was measured by room temperature
electronics, the signal had to be transmitted from the cryostat to
the amplifier by coaxial cables. Therefore, in view of impedance
matching, this set-up works well only for noise generating devices
with resistances of order 50$\Omega$. The more recent experiments
employ on-chip noise detectors, either quantum point
contacts\cite{Fujisaka, Ensslin} or Josephson
junctions.\cite{Pekola, Pothier} A first suggestion to use
Josephson junctions as threshold noise detectors was made by
Tobiska and Nazarov\cite{Nazarov} in 2004, and since then various
aspects of this idea have been analyzed by several
authors.\cite{Pekola1, Heikkila, Ankerhold1, Brosco, Ankerhold2,
Sukhorukov, Hekking}

Two recent experiments\cite{Pekola,Pothier} have studied the noise
generated by a tunnel junction through measurements of the
switching rate of an on-chip Josephson junction out of the
zero-voltage state. The skewness of the noise can be extracted
from the asymmetry of the switching rate with respect to the
direction of the bias current. In the region of noise activated 
escape, relevant for the experiments, the switching of a 
Josephson junction noise detector has been investigated in two 
recent papers. The work by Ankerhold\cite{Ankerhold2} describes 
the dynamics of the Josephson junction in terms of a 
Fokker-Planck equation driven by external noise. An approximate 
analytical expression for the switching rate is obtained for the 
entire range of damping parameters. The subsequent work by 
Sokhorukov and Jordan\cite{Sukhorukov} employs a path integral 
formalism and accounts for the feedback of the noise detector on 
the noise generating device. The authors also derive 
asymptotically exact results for the switching rate in the weak 
noise limit, however, only for the cases of vanishing damping and 
strong overdamping. In these limiting cases the problem 
simplifies considerably, since the number of relevant state 
variables is halved. The experimentally significant parameter 
range is at intermediate damping.

The aim of the present work is to provide, for the region of
activated escape in the weak noise limit, an asymptotically exact 
solution for the switching rate of a Josephson junction in 
presence of a device that generates non-Gaussian noise. The 
mutual influence of the two mesoscopic devices, Josephson noise 
detector and noise generator, will fully be taken into account by 
treating them on an equal footing. Furthermore, the entire range 
a damping parameters of the Josephson junction will be covered. 

The article is organized as follows. Sec.~\ref{sec_OM} briefly 
reviews a simplified version, sufficient to the present purposes, 
of the path integral representation of nonequilibrium 
thermodynamics in terms of thermodynamically conjugate variables. 
This approach was introduced more than fifty years ago by Onsager 
and Machlup\cite{Onsager} for the linear range near equilibrium 
and was then extended to the nonlinear range by Grabert, Graham, 
and Green.\cite{GG,GGG} The method, which is based on the 
conventional concept of Gaussian fluctuations, will then be 
applied in Sec.~\ref{sec_JJ} to the thermal escape of a Josephson 
junction driven by Johnson-Nyquist noise. These two introductory 
sections will also serve to introduce the relevant notation. The 
model described in Sec.~\ref{sec_JJ} will then be extended in 
Sec.~\ref{sec_NG} to account for non-Gaussian noise generated by 
a nonlinear device. Finally, Sec.~\ref{sec_dis} discusses 
concrete results for the experimentally relevant range of 
parameters and presents our conclusions. Some more technical 
details are moved to appendices.

%%%%%%%%%%%%%%%%%%%%%%%%%%%%%%%%%%%%%%%%%%%%%%%%%
%%%%%%%%%%%%%%%%%%%%%%%%%%%%%%%%%%%%%%%%%%%%%%%%%
%%%%%%%%%%%%%%%%%%%%%%%%%%%%%%%%%%%%%%%%%%%%%%%%%
%%%%%%        OMGGG-Approach              %%%%%%%
%%%%%%%%%%%%%%%%%%%%%%%%%%%%%%%%%%%%%%%%%%%%%%%%%
%%%%%%%%%%%%%%%%%%%%%%%%%%%%%%%%%%%%%%%%%%%%%%%%%
%%%%%%%%%%%%%%%%%%%%%%%%%%%%%%%%%%%%%%%%%%%%%%%%%
\section{Path Integral Representation of Fluctuations in Nonlinear Irreversible Processes}
\label{sec_OM}

Einstein\cite{Einstein} and Onsager\cite{Onsager1} have related
the stochastic theory of spontaneous fluctuations about
equilibrium with the deterministic theory of irreversible
processes. Perhaps the most seminal expression of this relation
between irreversible processes and fluctuations is the path
integral representation for the transition probability between two
macroscopic states. This functional, which gives a generalization
of the Boltzmann probability distribution to the time domain, was
introduced by Onsager and Machlup\cite{Onsager} for the linear
range near equilibrium and extended to nonlinear processes by
Grabert, Graham, and Green.\cite{GG,GGG}

Originally, the theory was formulated for closed systems where the
entropy is the appropriate thermodynamic potential. Here we want
to apply the method to describe mesoscopic systems that exchange
energy with a cryostat. The modifications are, of course,
well-known. The entire closed mega-system is divided into the
system of interest and the heat bath at constant temperature $T$,
and the Helmholtz free energy $F$ becomes the relevant
thermodynamic potential to characterize the system of interest.
When the state of this system is described in terms of the state
variables $a=(a_1,\ldots ,a_{\rm N})$, the Onsager transport
equations take the form
\begin{equation}\label{ote}
\dot a_{\rm I} = f_{\rm I} = \sum_{\rm J} L_{\rm IJ}\lambda_{\rm
J}\, ,
\end{equation}
where the $L_{\rm IJ}$ are the Onsager transport coefficients,
while the
\begin{equation}\label{otf}
\lambda_{\rm I}=-\frac{1}{T}\frac{\partial F}{\partial a_{\rm I}}
\end{equation}
are the thermodynamic forces. The transport equations are
nonlinear, if the thermodynamic forces are nonlinear functions of
the state variables $a$ or if the transport coefficients $L_{\rm
IJ}$ depend on the state variables. As will be seen below, for the
problem addressed here, the state dependence of the transport
coefficients is not relevant, and it will therefore be assumed
that the $L_{\rm IJ}$ are constant, they may depend on temperature
and other external parameters though. This simplifies the general
theory treated in Refs.~\onlinecite{GG,GGG} quite
considerably.\cite{remark1}

The state variables $a_{\rm I}$ can be chosen to be either even or
odd under time reversal
\begin{equation}\label{timeref}
   \tilde a_{\rm I} = \varepsilon_{\rm I} a_{\rm I}\ , \varepsilon_{\rm I} = \left\{
   \begin{array}{cc}
     \phantom{-}1, & \hbox{for even variables} \\
     &\\
     -1, & \hbox{for odd variables} \, .\\
   \end{array} \right.\,
\end{equation}
The Helmholtz free energy is an even variable
\begin{equation}\label{freeenergytime}
   F(\tilde a)= F(a)\, ,
\end{equation}
and the transport coefficients obey the reciprocal relations
\begin{equation}\label{reciprocal}
   L_{\rm IJ}(\tilde a) = \varepsilon_{\rm I} \varepsilon_{\rm J} L_{\rm IJ}(a)\, .
\end{equation}
The matrix $L_{\rm IJ}$ may be split into a symmetric part
\begin{equation}\label{dij}
   D_{\rm IJ}=\frac{1}{2}\left[L_{\rm IJ}+L_{\rm JI} \right]
\end{equation}
and an antisymmetric part
\begin{equation}\label{aij}
   A_{\rm IJ}=\frac{1}{2}\left[L_{\rm IJ}-L_{\rm JI} \right]\, .
\end{equation}
This implies a decomposition of the deterministic fluxes $f_{\rm
I}$ into a reversible drift
\begin{equation}\label{reversibledrift}
   r_{\rm I}=\sum_{\rm J} A_{\rm IJ}\lambda_{\rm J}\, ,
\end{equation}
with the symmetry $r_{\rm I}(\tilde a) = -\varepsilon_{\rm I}
r_{\rm I}(a)$, and an irreversible drift
\begin{equation}\label{irreversibledrift}
   d_{\rm I}=\sum_{\rm J} D_{\rm IJ}\lambda_{\rm J}\, ,
\end{equation}
with $d_{\rm I}(\tilde a) = \varepsilon_{\rm I} d_{\rm I}(a)$.
Only the irreversible drift contributes to the time rate of change
of the free energy
\begin{equation}\label{dotfreeenergy}
\dot F = -T\sum_{\rm I}\lambda_{\rm I}\dot a_{\rm I}=-T\sum_{\rm
I}\lambda_{\rm I} d_{\rm I}= -T \sum_{\rm I,J} D_{\rm
IJ}\lambda_{\rm I} \lambda_{\rm J}\, .
\end{equation}

Often, and in particular for the systems treated below, some of
the state variables do not couple directly to microscopic degrees
of freedom, and their fluxes are purely reversible. We then chose
the set of state variables $a$ so that the first n variables
$(a_1,\ldots,a_{\alpha},\ldots,a_{\rm n})$ are those with purely
reversible fluxes
\begin{equation}\label{reversiblevar}
   \dot a_{\alpha}=f_{\alpha}=r_{\alpha}\, .
\end{equation}
These variables will be distinguished by Greek indices
$\alpha,\beta$, while the remaining variables $(a_{\rm
n+1},\ldots,a_{\rm i}, \ldots, a_{\rm N})$ with partly
irreversible fluxes will be marked by small roman indices i,j. As
previously, large roman indices I,J run through the complete set
from 1 to N. Since the first n transport equations of the set
(\ref{ote}) take the form (\ref{reversiblevar}), the symmetric
parts of some of the transport coefficients vanish
\begin{equation}\label{dzero}
   D_{\alpha,\beta}=D_{\rm \alpha,i}=D_{\rm i,\alpha}=0\, .
\end{equation}

In the stochastic theory of irreversible processes the
irreversible drift is intimately connected with spontaneous
fluctuations about the deterministic motion.\cite{Einstein} These
fluctuations can be accounted for by random contributions
$\eta_{\rm I}$ to the thermodynamic forces $\lambda_{\rm I}$.
Following the approach by Grabert, Graham, and Green,\cite{GG,GGG}
the stochastic theory can be described in terms of a
Hamiltonian\cite{remark2}
\begin{equation}\label{hamiltonian}
   H(a,\eta)=\frac{1}{2}\sum_{\rm I,J}D_{\rm IJ}\eta_{\rm I}\eta_{\rm J} +\sum_{\rm I}
   f_{\rm I}(a)\eta_{\rm I} \, ,
\end{equation}
which implies equations of motion of canonical form
\begin{eqnarray}
\nonumber
 \dot a_{\rm I} &=& \frac{\partial H}{\partial \eta_{\rm I}}
 = f_{\rm I} + \sum_{\rm J} D_{\rm IJ}\eta_{\rm J}\\
 \dot \eta_{\rm I} &=& - \frac{\partial H}{\partial a_{\rm I}}
 = -\sum_{\rm J} \frac{\partial f_{\rm J}}{\partial
 a_{\rm I}} \eta_{\rm J}\, .
 \label{canonical}
\end{eqnarray}
Note that the deterministic transport equations (\ref{ote}) are
special solutions of (\ref{canonical}) with $\eta_{\rm I}=0$.

The canonical equations can be interpreted as Euler-Lagrange
equations and constraints (for the purely reversible fluxes) of an
action principle. The action determines the probability of a
fluctuation path, and the transition probability from an initial
state $a(0)=a$ to a final state $a(t)=a^{\prime}$ may be written
as a path integral
\begin{equation}\label{pi}
   p_t(a^{\prime}\vert a)
   =\int D[a,\eta] \exp{\left\{-\frac{1}{2k_B} A[a,\eta]\right\}}\, ,
\end{equation}
with the action functional
\begin{equation}\label{action}
   A[a,\eta]=\int_0^t ds \sum_{\rm I} \eta_{\rm I}\dot a_{\rm I} - H(a,\eta)\, .
\end{equation}
Since in view of Eq.~(\ref{dzero}) the Hamiltonian
(\ref{hamiltonian}) has quadratic terms for the $\eta_{\rm i}$
only, the action functional is linear in the $\eta_{\alpha}$ which
act as Lagrange parameters enforcing the constraints
(\ref{reversiblevar}). The $\eta_{\rm i}$, on the other hand, are
random forces describing fluctuations away from the deterministic
motion. The Hamiltonian is quadratic in the $\eta_{\rm i}$ because
of the underlying assumption of Gaussian fluctuations. For
mesoscopic systems this assumption may not be sufficient and an
appropriate extension of the approach to incorporate non-Gaussian
noise will be given in Sec.~\ref{sec_NG}.

%%%%%%%%%%%%%%%%%%%%%%%%%%%%%%%%%%%%%%%
%%%%%%%%%%%%%%%%%%%%%%%%%%%%%%%%%%%%%%%
%%%%%      SECTION JJ            %%%%%%
%%%%%%%%%%%%%%%%%%%%%%%%%%%%%%%%%%%%%%%
%%%%%%%%%%%%%%%%%%%%%%%%%%%%%%%%%%%%%%%
\section{Thermal Escape of a Josephson Junction From the Zero-Voltage State}
\label{sec_JJ}

In this section the thermally activated escape of a Josephson
junction form the zero-voltage state\cite{Ambegaokar} is reviewed
utilizing the approach outlined in the previous section.
%%%%%%%%%%%%%%%%%%%%%%%%%%%%%%%%%%%%%%%
%%%%%%%%%%%%%%%%%%%%%%%%%%%%%%%%%%%%%%%
\subsection{Transport Equations of a Biased Josephson Junction}
\label{sec_JJA}

\begin{figure}
\vspace{0.3cm}
\begin{center}
\includegraphics[scale=0.5]{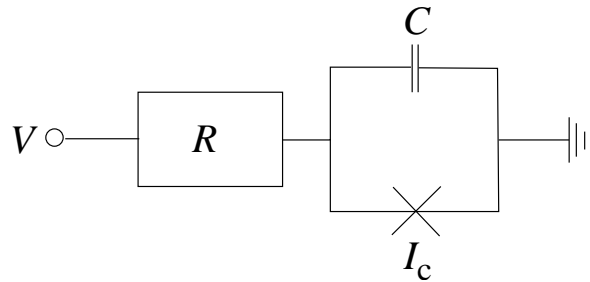}
\caption{\label{circuit}
Circuit diagram of a Josephson junction with critical current
$I_c$ and capacitance $C$ biased by a voltage source $V$ via a
resistor $R$.}
\end{center}
\end{figure}

The state variables of the Josephson junction are the charge $Q$
on the junction capacitance $C$ and the phase difference $\varphi$
between the order parameters of the superconductors on either side
of the tunnel barrier. The time rate of change of the phase is
related to the voltage $V_J$ across the Josephson junction by the
Josephson relation\cite{Josephson}
\begin{equation}\label{jrelation}
   V_J=\frac{\hbar}{2e}\dot \varphi\, .
\end{equation}
When a voltage $V$ is applied to a Josephson junction in series
with an Ohmic resistor $R$, as depicted in the circuit diagram
Fig.~\ref{circuit}, the electrical current $I$ flowing through
resistor and junction reads
%%%%%%%%%%%%%%%%%%%%%%%%%%%%%%%%%%%%%%%
%%%%%%%%%%%%%%%%%%%%%%%%%%%%%%%%%%%%%%%
%%%%%      FIGURE    1          %%%%%%
%%%%%%%%%%%%%%%%%%%%%%%%%%%%%%%%%%%%%%%
%%%%%%%%%%%%%%%%%%%%%%%%%%%%%%%%%%%%%%%
\begin{equation}\label{current}
   I=\frac{1}{R}\left(V-V_J\right)=\dot Q + I_c \sin(\varphi)\, ,
\end{equation}
where the second equality follows with the help of Josephson's
relation $I_s=I_c \sin(\varphi)$ for the supercurrent $I_s$ across
the junction. Combining Eqs.~(\ref{jrelation}), (\ref{current})
with $V_J=Q/C$, we readily find the deterministic equations of
motion
\begin{eqnarray}
\nonumber
 \dot\varphi &=& \frac{2e}{\hbar}\frac{Q}{C}\\
 \dot Q &=&  \frac{1}{R}\left(V-\frac{Q}{C}\right)-I_c
 \sin (\varphi )\, .
\label{equationom}
\end{eqnarray}
Clearly, $\varphi$ is a variable with purely reversible flux.\\

Let us introduce the free energy\begin{equation}\label{freeenergy}
   F(Q,\varphi)=F_0(T,V)+\frac{Q^2}{2C}-\frac{\hbar}{2e}\left[I_c\cos(\varphi)
+ \frac{V}{R}\varphi\right],
\end{equation}
and the thermodynamic forces
\begin{eqnarray}
\nonumber \lambda_{\varphi}&=& -\frac{1}{T}\frac{\partial
F}{\partial \varphi}
=-\frac{1}{T}\frac{\hbar}{2e}I_c\left[\sin(\varphi)-\frac{V}{R}\right]
\\
\nonumber\\
\lambda_Q &=& -\frac{1}{T}\frac{\partial F}{\partial Q}
=-\frac{1}{T}\frac{Q}{C} \, .\label{tforces}
\end{eqnarray}
The equations of motion (\ref{equationom}) can be then written in
Onsager form
\begin{equation}\label{onsagereom}
   \left(%
\begin{array}{c}
 \dot \varphi\\
 \\
 \dot Q \\
\end{array}%
\right)
=\left(%
\begin{array}{cc}
0 &-\frac{2eT}{\hbar}  \\
  &  \\
\frac{2eT}{\hbar}  & \frac{T}{R}\\
\end{array}%
\right)
\left(%
\begin{array}{cc}
\lambda_{\varphi} \\
   \\
 \lambda_Q \\
\end{array}%
\right)\, .
\end{equation}
Following the approach outlined in the previous section, and
denoting the conjugate variables to $(\varphi,Q)$ by
$(\mu,\lambda)$, the Hamiltonian of the system is found to read
\begin{eqnarray}
   H(\varphi,Q,\eta,\lambda)&=& \frac{T}{2R}\lambda^2 +
   \frac{1}{R}\left(V- \frac{Q}{C}\right) \lambda
   \nonumber\\
   &&-I_c\sin(\varphi)\lambda + \frac{2e}{\hbar}\frac{Q}{C}\mu\, ,
   \label{hamiltonianjj}
\end{eqnarray}
leading to the canonical equations
\begin{eqnarray}
\nonumber
 \dot\varphi &=& \frac{\partial H}{\partial \mu} = \frac{2e}{\hbar}\frac{Q}{C}\\
\nonumber   &&  \\
\nonumber \dot Q &=&  \frac{\partial H}{\partial \lambda} =
\frac{1}{R}\left(V- \frac{Q}{C}\right)
   -I_c\sin(\varphi)+ \frac{T}{R}\lambda\\
\nonumber   &&  \\
\nonumber
\dot \mu &=& -\frac{\partial H}{\partial \varphi} = I_c\cos(\varphi)\lambda \\
\nonumber   && \\
\dot \lambda &=& -\frac{\partial H}{\partial Q} =
\frac{1}{RC}\lambda - \frac{2e}{\hbar}\frac{\mu}{C}\,
.\label{canonicaljj}
\end{eqnarray}
While the purely reversible flux $\dot\varphi$ remains unchanged
in the stochastic theory, the flux $\dot Q$ is now supplemented by
a current $(T/R)\lambda$ describing Gaussian Johnson-Nyquist noise
from the Ohmic resistor.
%%%%%%%%%%%%%%%%%%%%%%%%%%%%%%%%%%%%%%%
%%%%%%%%%%%%%%%%%%%%%%%%%%%%%%%%%%%%%%%
\subsection{Decay of the Zero-Voltage State}
\label{sec_JJB}

As is apparent from Eq.~(\ref{freeenergy}), the Josephson junction
moves in the effective ``tilted washboard" potential
\begin{equation}\label{potential}
   U(\varphi)=-\frac{\hbar}{2e}\left[ I_c\cos(\varphi)+\frac{V}{R}\varphi
   \right]\, .
\end{equation}
It is convenient to introduce the dimensionless bias current
\begin{equation}\label{ess}
   s=\frac{V}{RI_c}\, .
\end{equation}
Then, for $0<s<1$, the potential has extrema in the phase interval
$[0,2\pi]$ at
\begin{equation}\label{well}
   \varphi_{\rm well, top}=\arcsin(s)=\frac{\pi}{2}\mp \delta\, ,
\end{equation}
where for $1-s\ll 1$
\begin{equation}\label{delta}
   \delta\approx\sqrt{2(1-s)}\, .
\end{equation}
When the Josephson junction is trapped in the state $\varphi_{\rm
well}=\frac{\pi}{2}-\delta$, the average voltage $V_J$ across the
junction vanishes. However, this zero-voltage state is
metastable, since the well is only a local minimum of the
potential (\ref{potential}).  To escape from the well, the
junction needs to be thermally activated to the barrier top at
$\varphi_{\rm top}=\frac{\pi}{2}+ \delta$. This process will be
observed with large probability, when the barrier height is
small, which is the case when the dimensionless bias current $s$
is close to 1. We shall not discuss here escape by macroscopic
quantum tunneling,\cite{CL} which occurs at very low temperatures.

The decay rate follows from the transition probability from
$\varphi_{\rm well}$ to $\varphi_{\rm top}$ as governed by the
path integral (\ref{pi}). The dominant contribution to the
functional integral comes from the minimal action path satisfying
the canonical equations (\ref{canonicaljj}). Let us first
consider the reverse process, the relaxation from the barrier top
$\varphi_{\rm top}$ to the well minimum $\varphi_{\rm well}$. In
this case the most probable path is the deterministic path, that
is a solution of the evolution equations (\ref{canonicaljj}) with
$\mu=\lambda=0$. The two remaining equations of motion can be
combined to read
\begin{equation}
\label{relaxation}
   \frac{\hbar}{2e}C\ddot\varphi +\frac{\hbar}{2e}\frac{1}{R}\dot\varphi
   +I_c\sin(\varphi)=\frac{V}{R}\, .
\end{equation}\\
There is a solution\cite{remark3} $\varphi_{\rm relax}(t)$ of
(\ref{relaxation}) satisfying
\begin{equation}\label{relaxsolution}
   \varphi_{\rm relax}(-\infty)=\varphi_{\rm top},\ \varphi_{\rm relax}(+\infty)=\varphi_{\rm well}\, ,
\end{equation}
which describes the relaxation from the barrier top to the well
bottom. Since $\mu$ and $\lambda$ vanish, this deterministic
trajectory has vanishing action (\ref{action}).

The minimal action trajectory for thermally activated escape from
the zero-voltage state $\varphi_{\rm well}$ is a solution
$\varphi_{\rm esc}(t)$ of the canonical equations
(\ref{canonicaljj}) with
\begin{equation}\label{activation}
   \varphi_{\rm esc}(-\infty)=\varphi_{\rm well},\
   \varphi_{\rm esc}(+\infty)=\varphi_{\rm
   top}\, .
\end{equation}
The first two of the canonical equations (\ref{canonicaljj})
combine to give
\begin{equation}
\label{canonicalcombi1}
   \frac{\hbar}{2e}C\ddot\varphi +\frac{\hbar}{2e}\frac{1}{R}\dot\varphi
   +I_c\sin(\varphi)=\frac{V+T\lambda}{R}\, .
\end{equation}\\
Now, the ansatz $\varphi_{\rm esc}(t)=\varphi_{\rm relax}(-t)$
satisfies the boundary conditions (\ref{activation}) and also the
evolution equation (\ref{canonicalcombi1}) provided
\begin{equation}\label{condition}
   \lambda_{\rm esc}(t)=-\frac{\hbar}{eT}\dot \varphi_{\rm relax}(-t)=\frac{\hbar}{eT}\dot \varphi_{\rm esc}(t)\, ,
\end{equation}
where we have used the fact that $\varphi_{\rm relax}(t)$ is a
solution of Eq.~(\ref{relaxation}) with boundary conditions
(\ref{relaxsolution}), and that $\dot\varphi_{\rm esc}(t)=-\dot
\varphi_{\rm relax}(-t)$, $\ddot \varphi_{\rm esc}(t)=\ddot
\varphi_{\rm relax}(-t)$. The last equation of the set
(\ref{canonicaljj}) then gives
\begin{eqnarray}
\nonumber
   \mu_{\rm esc}(t)&=&-\frac{\hbar
   C}{2e}\left[\dot\lambda_{\rm esc}(t)-\frac{1}{RC}\lambda_{\rm esc}(t)\right]\\
   &&\nonumber\\
   &=&-\frac{\hbar}{eT}\left[\frac{\hbar}{2e}C\ddot \varphi_{\rm relax}(-t)
   +\frac{\hbar}{2e}\frac{1}{R}\dot \varphi_{\rm relax}(-t) \right]
   \nonumber\\
   &&\nonumber\\
   &=&-\frac{\hbar}{eT} \left[ \frac{V}{R} - I_c
   \sin\left(\varphi_{\rm relax}(-t)\right)\right]\, ,
   \label{musolution}
\end{eqnarray}
where we have again used the equation of motion (\ref{relaxation})
satisfied by $\varphi_{\rm relax}(t)$ to derive the last line.
Now, Eqs.~(\ref{condition}) and (\ref{musolution}) combine to give
\begin{equation}\label{condition2}
   \dot \mu_{\rm esc}(t) = I_c \cos\left(\varphi_{\rm
   esc}(t)\right)\lambda_{\rm esc}(t)\, ,
\end{equation}
so that the remaining equation of the canonical set of equations
(\ref{canonicaljj}) is also satisfied, and the ansatz
$\varphi_{\rm esc}(t)=\varphi_{\rm relax}(-t)$ gives indeed the
minimal action escape path.

To determine the action (\ref{action}) of the escape path, we
first note that the Hamiltonian (\ref{hamiltonianjj}), which is
conserved along a solution of the canonical equations, vanishes on
the escape path, since $\lambda_{\rm esc}(\pm\infty)=\mu_{\rm
esc}(\pm\infty)=0$, as can be inferred from Eqs.~(\ref{condition})
and (\ref{musolution}). Thus
\begin{eqnarray}\label{actionesc}
\nonumber
   A_{\rm esc}&=&\int_{-\infty}^{\infty} dt\left[ \lambda_{\rm esc}(t)\dot Q_{\rm esc}(t)
   +\mu_{\rm esc}(t)\dot \varphi_{\rm esc}(t)\right]\\
\nonumber\\
\nonumber
   &=&\int_{-\infty}^{\infty} dt\bigg\{\frac{\hbar}{eT}\dot \varphi_{\rm esc}(t)
   \frac{\hbar}{2e}C\ddot\varphi_{\rm esc}(t)  \\
\nonumber\\
&&-\frac{\hbar}{eT} \left[ \frac{V}{R} - I_c
   \sin\left(\varphi_{\rm esc}(t)\right)\right]
   \dot \varphi_{\rm esc}(t)\bigg\}\, ,
\end{eqnarray}
where we have used the first of the canonical equations
(\ref{canonicaljj}) as well as Eqs.~(\ref{condition}) and
(\ref{musolution}) to express $\dot Q_{\rm esc}(t)$, $\lambda_{\rm
esc}(t)$, and $\mu_{\rm esc}(t)$ in terms of $\varphi_{\rm
esc}(t)$. The result (\ref{actionesc}) may now be transformed to
read
\begin{eqnarray}\label{actionesc2}
\nonumber
   A_{\rm esc}&=&\frac{1}{T}\int_{-\infty}^{\infty} dt
   \left\{ \left(\frac{\hbar}{2e} \right)^2
   C\frac{\partial}{\partial t}\dot\varphi_{\rm esc}^2
   +2 U^{\prime}(\varphi_{\rm esc}) \, \dot \varphi_{\rm esc}\right\} \\
\nonumber\\
  &=& \frac{2}{T}\left[ U(\varphi_{\rm top})
-U(\varphi_{\rm
  well})\right]\, ,
\end{eqnarray}
where the last expression follows from the boundary conditions
(\ref{activation}) obeyed by $\varphi_{\rm esc}(t)$ for
$t\to\pm\infty$.

The rate of escape $\Gamma$ from the metastable well may be
written as
\begin{equation}\label{rate}
   \Gamma = f\,{\rm e}^{-B}\, ,
\end{equation}
where the exponential factor $B$ is determined by the action of
the most probable escape path $\varphi_{\rm esc}$ of the path
integral.\cite{schulman} Introducing the barrier height
\begin{equation}\label{barrierheight}
 \Delta U = U(\varphi_{\rm top}) -U(\varphi_{\rm well})\, ,
\end{equation}
we obtain from Eqs.~(\ref{pi}) and (\ref{actionesc2}) for the
exponential factor
\begin{equation}\label{arrhenius}
   B= \frac{\Delta U}{k_BT}\, ,
\end{equation}
which is just the standard Arrhenius factor for thermally
activated decay. The pre-exponential factor $f$ requires an
analysis of the fluctuations about the minimal action path and
will not be addressed here.

%%%%%%%%%%%%%%%%%%%%%%%%%%%%%%%%%%%%%%%
%%%%%%%%%%%%%%%%%%%%%%%%%%%%%%%%%%%%%%%
%%%%%       SECTION NG           %%%%%%
%%%%%%%%%%%%%%%%%%%%%%%%%%%%%%%%%%%%%%%
%%%%%%%%%%%%%%%%%%%%%%%%%%%%%%%%%%%%%%%
\section{Josephson Junction Driven by Non-Gaussian Noise}
\label{sec_NG}

So far we have studied a biased Josephson junction driven by
Gaussian thermal noise. We now address the question how the rate
of escape $\Gamma$ from the zero-voltage state is modified by the
presence of non-Gaussian noise. To be specific, we shall consider
the shot noise generated by a normal state tunnel junction, since
this case has been examined in recent
experiments.\cite{Pekola,Pothier} However, the theory likewise
applies to other noise generating devices with short noise
correlation times.
%%%%%%%%%%%%%%%%%%%%%%%%%%%%%%%%%%%%%%%
%%%%%%%%%%%%%%%%%%%%%%%%%%%%%%%%%%%%%%%
\subsection{Hamiltonian for Non-Gaussian Noise}
\label{sec_NGA}

\begin{figure}
\vspace{0.3cm}
\begin{center}
\includegraphics[scale=0.4]{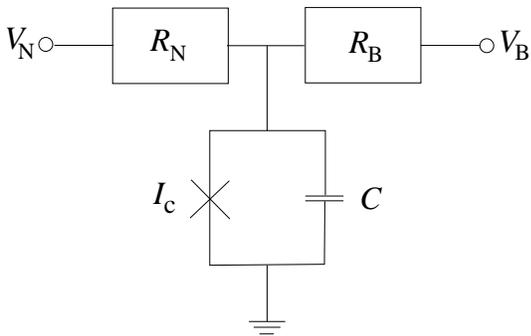}
\caption{\label{circuit2}
Circuit diagram of a Josephson junction with critical current
$I_c$ and capacitance $C$ biased in a twofold way. The branch to
the right puts an Ohmic resistor $R_B$ is series with the junction
and is biased by the voltage $V_B$. The branch to the left is
biased by a voltage $V_N$ and $R_N$ is a noise generating
nonlinear element, specifically a normal state tunnel junction
with tunnelling resistance $R_N$.}
\end{center}
\end{figure}

Let us consider a Josephson junction with capacitance $C$ and
critical current $I_c$ driven by two noise sources, see
Fig.~\ref{circuit2}. A bias voltage $V_B$ is applied to one branch
with an Ohmic resistor $R_B$ in series with the junction. This
part of the set-up corresponds to the model treated in the
previous section. A second voltage $V_N$ is applied to another
branch with a tunnel junction of resistance $R_N$ again in series
with the Josephson junction. Experimental set-ups are typically
more sophisticated, but the circuit diagram in
Fig.~\ref{circuit2} captures the essentials of a Josephson
junction on-chip noise detector.
%%%%%%%%%%%%%%%%%%%%%%%%%%%%%%%%%%%%%%%
%%%%%%%%%%%%%%%%%%%%%%%%%%%%%%%%%%%%%%%
%%%%%      FIGURE    2          %%%%%%
%%%%%%%%%%%%%%%%%%%%%%%%%%%%%%%%%%%%%%%
%%%%%%%%%%%%%%%%%%%%%%%%%%%%%%%%%%%%%%%
The current $I$ flowing through the Josephson junction is given by
\begin{equation}
\label{currentng}
   I=\frac{V_B-V_J}{R_B}+
   \frac{V_N-V_J}{R_N}=\dot Q + I_c \sin(\varphi)\, .
\end{equation}
Proceeding as in Sec.~\ref{sec_JJ}, one readily obtains the
deterministic equations of motion
\begin{eqnarray}
 \dot\varphi &=& \frac{2e}{\hbar}\frac{Q}{C}\\
 \dot Q &=&  \frac{1}{R_B}\left(V_B-\frac{Q}{C}\right)
 + \frac{1}{R_N}\left(V_N-\frac{Q}{C}\right)-I_c
 \sin (\varphi )\, .\nonumber
\label{equationng}
\end{eqnarray}
Since the flux $\dot \varphi$ is purely reversible, the
Hamiltonian $H(\varphi,Q,\eta,\lambda)$ will depend on the
conjugate variable $\eta$ only linearly, while the dependence on
$\lambda$ comprises linear and nonlinear terms. In contrast to the
case studied in the previous section, the nonlinear terms in
$\lambda$ will not be just quadratic, since the noise generated by
the normal state tunnel junction is non-Gaussian. As the voltage
$V_N^{\prime}=V_N-V_J$ across the tunnel junctions grows relative
to $k_BT/e$, the noise generated by the tunnel junction crosses
over from Gaussian to Poissonian statistics. For the current $I_N$
through the tunnel junction one has\cite{reviews}
\begin{eqnarray}
\nonumber
  \left\langle I_N \right\rangle&=& \frac{V_N^{\prime}}{R_N} \\
 \nonumber
\\
 \left\langle \delta I_N(t)\,\delta I_N(t^{\prime}) \right\rangle &=&
 C_2\,\delta(t-t^{\prime}) \\
 \nonumber\\
 \nonumber
\left\langle \delta I_N(t)\,\delta I_N(t^{\prime})\,\delta
I_N(t^{\prime\prime}) \right\rangle &=&
C_3\,\delta(t-t^{\prime})\,\delta(t^{\prime}-t^{\prime\prime})\, ,
 \label{noise}
\end{eqnarray}
where $\delta I_N(t)= I_N -\left\langle I_N \right\rangle$ and
\begin{eqnarray}
\nonumber
 C_2 &=& \frac{eV_N^{\prime}}{R_N}
 \coth\left(\frac{eV_N^{\prime}}{2k_BT}\right)\\
 \nonumber\\
 C_3 &=& \frac{e^2V_N^{\prime}}{R_N}\, .
\end{eqnarray}
There are of course higher order noise cumulants, but, as we
shall see, these are not important in the region of noise 
activated switching of the Josephson noise detector. The skewness 
of the noise described by $C_3$ leads to a cubic term in 
$\lambda$. Neglecting terms of fourth order, the Hamiltonian 
takes the form\cite{else}
\begin{eqnarray}
\nonumber
 &&  H(\varphi,Q,\eta,\lambda)=
   \frac{T}{2R_B}\lambda^2 +
   \frac{1}{R_B}\left(V_B- \frac{Q}{C}\right) \lambda \\
   \nonumber\\
   \nonumber
  &&+ \frac{e\left(V_N-\frac{Q}{C}\right)}{4k_BR_N}
    \coth\left[\frac{e\left(V_N-\frac{Q}{C}\right)}{2k_BT}
    \right]\lambda^2\\
   \nonumber\\
  &&+ \frac{1}{R_N}\left(V_N-\frac{Q}{C}\right)
   \left[\lambda+ \frac{1}{24}\left(\frac{e}{k_B}\right)^2\lambda^3\right] \nonumber \\
   \nonumber\\
  &&-I_c\sin(\varphi)\lambda + \frac{2e}{\hbar}\frac{Q}{C}\mu +\ {\cal O}\left(\lambda^4\right)\, .
   \label{hamiltonianng}
\end{eqnarray}
An expansion of the Hamiltonian in powers of $\lambda$ is
justified, provided the dimensionless quantity $e\lambda/k_B\ll
1$. As discussed in detail in App.~\ref{appa}, the size of the
random forces  $\lambda$ causing the escape is proportional to the
size of the fluctuations of the voltage $V_J$ across the Josephson
junction, and $e\lambda/k_B$ is in fact very small, if the decay
of the zero-voltage state occurs in the region of noise activated 
escape. Since $V_J=Q/C$ and $\lambda$ are effectively 
proportional to each other, it does not make sense to keep higher 
order terms in $Q/C$, rather, the two small parameters, 
$e\lambda/k_B$ and $Q/CV_N$, should be treated on an equal 
footing. Hence, the term in the second line of 
Eq.~(\ref{hamiltonianng}), which is already of second order in 
$\lambda$, can be expanded to first order in $Q/CV_N$. Likewise 
the $Q/CV_N$ dependence of the term of order $\lambda^3$ can be 
dropped. We then find
\begin{equation}\label{hamilng}
   H(\varphi,Q,\eta,\lambda)=H_2(\varphi,Q,\eta,\lambda)
   +H_3(\varphi,Q,\eta,\lambda)\, ,
\end{equation}
where
\begin{eqnarray}\label{hamilng2}
  \nonumber
   H_2(\varphi,Q,\eta,\lambda)&=&
   \left(\frac{T}{2R_B} + \frac{C_{2,N}}{4k_B}\right)\lambda^2
   \\\nonumber\\
   &+& \left( I_{\rm bias}- \frac{1}{R_{\vert\vert}} \frac{Q}{C}\right)
   \lambda \nonumber \\
   \nonumber\\
  &-& I_c\sin(\varphi)\lambda + \frac{2e}{\hbar}\frac{Q}{C}\mu\, ,
\end{eqnarray}
describes Gaussian noise. Here we have introduced the bias
current\cite{remark4}
\begin{equation}\label{bias}
   I_{\rm bias}=\frac{V_B}{R_B}+\frac{V_N}{R_N}\, ,
\end{equation}
the second noise cumulant
\begin{equation}\label{CN}
   C_{2,N}=\frac{eV_N}{R_N}
    \coth\left(\frac{eV_N}{2k_BT}\right)\, ,
\end{equation}
and the parallel resistance
\begin{equation}\label{parallelr}
  \frac{1}{R_{\vert\vert}}=\frac{1}{R_{B}} + \frac{1}{R_{N}}\, .
\end{equation}
The term
\begin{eqnarray}\label{hamilng3}
\nonumber
   H_3(\varphi,Q,\eta,\lambda)&=&\frac{1}{24k_B^2}
   C_{3,N} \lambda^3\\
   &&\nonumber\\
  &-& \frac{1}{4k_B}\frac{\partial C_{2,N}}{\partial V_N}\frac{Q}{C}\lambda^2
\end{eqnarray}
with the third noise cumulant
\begin{equation}\label{c3n}
   C_{3,N}=e^2\frac{V_N}{R_N}
\end{equation}
includes the leading order effects of non-Gaussian noise.
%%%%%%%%%%%%%%%%%%%%%%%%%%%%%%%%%%%%%%%
%%%%%%%%%%%%%%%%%%%%%%%%%%%%%%%%%%%%%%%
\subsection{Minimal Action Escape Path in the Nearly Gaussian Regime}
\label{sec_NGB}
In the range of parameters studied here, the third order
Hamiltonian (\ref{hamilng3}) will describe weak corrections to
the dynamics governed by the Hamiltonian (\ref{hamilng2}). In
fact, this latter Hamiltonian is precisely of the form of the
Hamiltonian (\ref{hamiltonianjj}) studied in Sec.~\ref{sec_JJ} for a
Josephson junction in parallel with on Ohmic conductor, provided we
replace $R$ by the parallel resistance $R_{\vert\vert}$, the current
$V/R$ by the proper bias current $I_{\rm bias}$, and $T$
by the effective temperature
\begin{eqnarray}\label{teff}
   T_{\rm eff}&=&R_{\vert\vert}\left[\frac{T}{R_B} + \frac{C_{2,N}}{2k_B} \right]
   \nonumber\\\nonumber\\
   &=&R_{\vert\vert}\left[\frac{T}{R_B} + \frac{eV_N}{2k_BR_N}
    \coth\left(\frac{eV_N}{2k_BT}\right) \right]\, .
\end{eqnarray}
For $eV_N\ll k_BT$ the tunnel junction generates approximately
Johnson-Nyquist noise and the effective temperature coincides with
the cryostat temperature. On the other hand, for $eV_N\gg k_BT$,
the tunnel junction is a source of shot noise with a noise power
proportional to $V_N$. The Josephson junction reacts to the
additional Gaussian noise in the same way as to an elevated
temperature.\cite{Huard} Approximate expressions for $T_{\rm eff}$
have been presented previously.\cite{Ankerhold2,Sukhorukov}
Experimentally, $T_{\rm eff}$ can be substantially larger than
$T$.

The rate of escape $\Gamma$ from the zero-voltage state of the
Josephson junction will again be of the form (\ref{rate}), where
the exponent $B$ now takes the form
\begin{equation}\label{rateng}
   B=B_2+B_3
\end{equation}
with
\begin{equation}\label{b2}
   B_2=\frac{\Delta U}{k_BT_{\rm eff}}\, .
\end{equation}
The exponential factor $B_2$ is determined by the action of the
approximate escape path $\varphi_2(t)$ that solves the canonical
equations of motion resulting from the second order Hamiltonian
(\ref{hamilng2}). The second cumulant (\ref{CN}) of the noise
generated by the normal state tunnel junction is taken into
account in terms of the effective temperature $T_{\rm eff}$. To
include the effects of the third cumulant $C_{3,N}$, one needs to
determine the deviation $\varphi_3(t)$ of the escape path from
$\varphi_2(t)$. To this purpose we start with the canonical
equations that follow from Eqs.~(\ref{hamilng}), (\ref{hamilng2})
and (\ref{hamilng3}). We find
\begin{eqnarray}
\nonumber
 \dot \varphi &=& \frac{\partial H}{\partial \mu} =
\frac{2e}{\hbar}\frac{Q}{C}  \\
 \nonumber
  \\
 \nonumber
 \dot Q &=& \frac{\partial H}{\partial \lambda} =
  I_{\rm bias} -\frac{1}{R_{\vert\vert}}\frac{Q}{C}
   -I_c\sin(\varphi)+ \frac{T_{\rm eff}}{R_{\vert\vert}}\lambda\\
\nonumber   &&  \\
&&+ \frac{1}{8k_B^2} \, C_{3,N} \lambda^2
   -\frac{1}{2k_B}\frac{\partial C_{2,N}}{\partial
V_N}\frac{Q}{C}\lambda \, ,\label{canonicalng1}
\end{eqnarray}
where we have made use of Eq.~(\ref{teff}), and
\begin{eqnarray}\nonumber
\dot \mu &=& -\frac{\partial H}{\partial \varphi} = I_c\cos(\varphi)\lambda \\
\nonumber   && \\
\dot \lambda &=& -\frac{\partial H}{\partial Q} =
\frac{1}{R_{\vert\vert}C}\lambda -
\frac{2e}{\hbar}\frac{\mu}{C}
\nonumber \\
&& \nonumber\\
 &&+ \frac{1}{4k_B}\frac{\partial C_{2,N}}{\partial V_N}
 \frac{\lambda^2}{C}   \, .
 \label{canonicalng2}
\end{eqnarray}
Now, the two differential equations (\ref{canonicalng1}) of first
order can be combined to one second order differential equation
\begin{equation}\label{eqmphi}
   \frac{\hbar}{2e}C\ddot\varphi +\frac{\hbar}{2e}\frac{1}{R_{\vert\vert}}\dot\varphi
   +I_c\sin(\varphi)=I_{\rm bias}+\frac{T_{\rm eff}}{R_{\vert\vert}}\lambda+I_3\, ,
\end{equation}
where\begin{equation}\label{i3}
   I_3= \frac{1}{8k_B^2}\, C_{3,N} \lambda^2
   -\frac{\hbar}{4ek_B}\frac{\partial C_{2,N}}{\partial V_N}\dot\varphi\lambda
\end{equation}
is the additional noise current arising from $H_3$. Likewise, the
Eqs.~(\ref{canonicalng2}) combine to give
\begin{equation}\label{eqmlambda}
   \frac{\hbar}{2e}C\ddot\lambda
   -\frac{\hbar}{2e}\frac{1}{R_{\vert\vert}}\dot\lambda
   +I_c\cos(\varphi)\lambda=I_3^{\prime}\lambda\, ,
\end{equation}
where
\begin{equation}\label{i3strich}
   I_3^{\prime}= \frac{\hbar}{4ek_B}\frac{\partial C_{2,N}}{\partial V_N}
 \dot\lambda\,
\end{equation}
again results from $H_3$.

We now make the ansatz
\begin{eqnarray}
\nonumber
 \varphi_{\rm esc}(t) &=& \varphi_2(t) + \varphi_3(t) \\
 \nonumber\\\
 \lambda_{\rm esc}(t) &=& \lambda_2(t) + \lambda_3(t)\, ,
 \label{ansatz}
\end{eqnarray}
where $\varphi_2(t)$ and $\lambda_2(t)$ are the solutions of
(\ref{eqmphi}) and (\ref{eqmlambda}) for $I_3=I_3^{\prime}=0$,
while $\varphi_3(t)$ and $\lambda_3(t)$ describe the modifications
of the path arising for finite $I_3$ and $I_3^{\prime}$. For
$I_3=0$, the equation of motion (\ref{eqmphi}) is of the form of
the evolution equation (\ref{canonicalcombi1}) studied in
Sec.~\ref{sec_JJ}, and we can proceed as there. Provided $s=I_{\rm
bias}/I_c < 1$, the potential $ U(\varphi)=-(\hbar/2e)I_c\left[
\cos(\varphi)+s\varphi \right]$ has a minimum $\varphi_{\rm well}$
and a maximum $\varphi_{\rm top}$ in the phase interval
$[0,2\pi]$. From a solution $\varphi_{\rm relax}(t)$ satisfying
\begin{equation}\label{relaxation2}
   \frac{\hbar}{2e}C\ddot\varphi +\frac{\hbar}{2e}\frac{1}{R_{\vert\vert}}\dot\varphi
   +I_c\sin(\varphi)=I_{\rm bias} ,
\end{equation}
and the boundary conditions (\ref{relaxsolution}), we obtain an
escape path satisfying the Eqs.~(\ref{eqmphi}) and
(\ref{eqmlambda}) for $I_3=I_3^{\prime}=0$ and the boundary
conditions
\begin{eqnarray}
\nonumber
 \varphi_2(-\infty)=\varphi_{\rm well} &,& \varphi_2(+\infty)=\varphi_{\rm top} \\
 \nonumber\\
 \lambda_2(-\infty)=0 &,& \lambda_2(+\infty)=0
\end{eqnarray}
by putting
\begin{eqnarray}
 \varphi_2(t) &=& \varphi_{\rm relax}(-t) \nonumber \\
 \nonumber\\
 \lambda_2(t) &=& \frac{\hbar}{eT_{\rm eff}}\dot\varphi_2(t)\, .
 \label{phi2}
\end{eqnarray}
Next, we insert the ansatz (\ref{ansatz}) into the evolution
equations (\ref{eqmphi}) and (\ref{eqmlambda}) and keep only terms
that are linear in the quantities $\varphi_3$, $\lambda_3$, $I_3$,
and $I_3^{\prime}$ which describe corrections to the Gaussian
case. Taking advantage of the equations of motion satisfied by
$\varphi_2$ and $\lambda_2$, we obtain
\begin{equation}\label{eqmphi3}
   \frac{\hbar}{2e}C\ddot\varphi_3
   +\frac{\hbar}{2e}\frac{1}{R_{\vert\vert}}\dot\varphi_3
   +I_c\cos(\varphi_2)\varphi_3=\frac{T_{\rm eff}}{R_{\vert\vert}}\lambda_3+I_3\, ,
\end{equation}
and
\begin{eqnarray}\label{eqmlambda3}
   \frac{\hbar}{2e}C\ddot\lambda_3
   -\frac{\hbar}{2e}\frac{1}{R_{\vert\vert}}\dot\lambda_3
   +I_c\cos(\varphi_2)\lambda_3
   \nonumber\\
   =\frac{\hbar}{eT_{\rm eff}}\dot \varphi_2\left[I_c\sin(\varphi_2)\varphi_3+I_3^{\prime}\right]\, ,
\end{eqnarray}
where $I_3$ and $I_3^{\prime}$ defined in (\ref{i3}) and
(\ref{i3strich}) are now evaluated with the leading order
solutions (\ref{phi2}). Hence
\begin{equation}\label{i30}
I_3=\frac{1}{2\left(k_BT_{\rm eff}\right)^2}
   \left(\frac{\hbar}{2e}\right)^2
   \left(     C_{3,N}  - 2k_BT_{\rm eff}\frac{\partial C_{2,N}}{\partial V_N}
   \right) \dot\varphi_2^2\, ,
\end{equation}
and
\begin{equation}\label{i3strich0}
   I_3^{\prime}= \frac{1}{k_BT_{\rm eff}}
   \left(\frac{\hbar}{2e}\right)^2
   \frac{\partial C_{2,N}}{\partial V_N}
\, \ddot\varphi_2\, .
\end{equation}
We shall see that an explicit solution of these evolution
equations is not required to determine the action.
%%%%%%%%%%%%%%%%%%%%%%%%%%%%%%%%%%%%%%%%%%
%%%%%%%%%%%%%%%%%%%%%%%%%%%%%%%%%%%%%%%%%%
\subsection{Action of Escape Path}
\label{sec_NGC} Since the Hamiltonian (\ref{hamilng}) vanishes
along the escape path, the action may be written
\begin{equation}\label{actionesc3}
\nonumber
   A_{\rm esc}=\int_{-\infty}^{\infty} dt\left[
    \lambda_{\rm esc}\dot Q_{\rm esc}
   -\dot\mu_{\rm esc} \varphi_{\rm esc}\right]\, ,
\end{equation}
where we have made a partial integration with respect to the first
line of Eq.~(\ref{actionesc}). From Eq.~(\ref{canonicalng1}), we
have $\dot Q_{\rm esc}=(\hbar/2e)C\ddot\varphi_{\rm esc}$, while
Eq.~(\ref{canonicalng2}) implies $\dot\mu_{\rm
esc}=I_c\cos(\varphi_{\rm esc}) \lambda_{\rm esc}$. Inserting this
as well as the ansatz (\ref{ansatz}) into the action
(\ref{actionesc3}), we find after disregarding terms of second
order in $\varphi_3$ and $\lambda_3$
\begin{equation}\label{actionesc4}
   A_{\rm esc}=A_2+A_3\, ,
\end{equation}
where
\begin{equation}\label{a2}
   A_2=\int_{-\infty}^{\infty} dt\left[
   \frac{\hbar}{2e}C\lambda_2\ddot\varphi_2
    - I_c \cos(\varphi_2)\lambda_2\varphi_2\right]\, ,
\end{equation}
and
\begin{eqnarray}
\label{a3}
   A_3&=&\int_{-\infty}^{\infty} dt\bigg[
    \frac{\hbar}{2e}C\left(\lambda_2\ddot\varphi_3
    +\lambda_3\ddot\varphi_2 \right)\\
    &&- I_c \cos(\varphi_2)\left(\lambda_2\varphi_3+\lambda_3\varphi_2\right)
    + I_c \sin(\varphi_2)\lambda_2\varphi_2\varphi_3
    \bigg]\, .\nonumber
\end{eqnarray}
Now, the deviations $\varphi_3$ and $\lambda_3$ from the path of
the Gaussian model are caused by the currents $I_3$ and
$I_3^{\prime}$ given in Eqs.~(\ref{i30}) and (\ref{i3strich0}).
These currents depend on the third noise cumulant $C_{3,N}$ and on
the derivative $\partial C_{2,N}/\partial V_N$ of the second
cumulant. The detailed evaluation of the action in App.~\ref{appb}
shows, that these two factors influence the action $A_3$ only in
the combination
\begin{equation}\label{c3a}
  \mathcal{C}_3=C_{3,N}- 3k_BT_{\rm eff}\frac{\partial C_{2,N}}{\partial
  V_N}\, .
\end{equation}
A corresponding reduction of the effective third cumulant was
already noted by Sokhurokov and Jordan\cite{Sukhorukov} for the
limiting cases of weak and strong damping. The second term in
Eq.~(\ref{c3a}) arises from the feedback of the Josephson junction
on the noise generating junction, which is a consequence of the
finite voltage $V_J$ that builds up during escape. Experiments are
usually done in the regime $eV_N \gg k_BT$, where
\begin{eqnarray}
\label{c3approx} \nonumber
   \mathcal{C}_3&\approx&
   C_{3,N}\left(1-
   \frac{3k_BT_{\rm eff}}{eV_N}\right)
   \\ \nonumber \\
   &\approx&
   C_{3,N}\left(1-
   \frac{3}{2}\frac{R_B+\frac{2k_BT}{eV_N}R_N}{R_B+R_N}\right)\, ,
\end{eqnarray}
so that the feedback becomes negligible for $R_N\gg R_B$. In the
opposite limit the feedback even changes the sign of
$\mathcal{C}_3$.

As shown in App.~\ref{appb}, repeated use of the equations of
motion satisfied by $\varphi_2$, $\lambda_2$, $\varphi_3$, and
$\lambda_3$ allows one to express $A_3$ entirely in terms of
$\varphi_2(t)$. By virtue of Eq.~(\ref{phi2}), $\varphi_2(t)$ is
time reversed to the deterministic trajectory $\varphi_{\rm
relax}(t)$ describing the relaxation from the barrier top.
Accordingly, the result (\ref{a3i}) in App.~\ref{appb} may be
written as
\begin{equation}\label{a3relax}
   A_3= -
\frac{2k_B}{\left(k_BT_{\rm eff}\right)^3}
\left(\frac{\hbar}{2e}\right)^3 \mathcal{C}_3 J \, .
\end{equation}
where
\begin{equation}\label{jrelax}
   J=-\frac{1}{6}\int_{-\infty}^{\infty} dt \, \dot\varphi_{\rm
   relax}^3(t)\ .
\end{equation}
Thus, the non-Gaussian correction to the rate exponent
(\ref{rateng}) reads
\begin{equation}\label{b3}
   B_3=
\frac{1}{\left(k_BT_{\rm eff}\right)^3}
\left(\frac{\hbar}{2e}\right)^3 \mathcal{C}_3 J \, .
\end{equation}
What remains to be determined is the quantity $J$, which describes
a property of the system in the absence of noise.

Let us introduce the energy function
\begin{equation}\label{energy}
   E(\varphi,\dot\varphi)=
   \frac{1}{2}\left(\frac{\hbar}{2e}\right)^2\! C\dot\varphi^2
   +U(\varphi)\, ,
\end{equation}
where $U(\varphi)$ is the potential (\ref{potential}) with $V/R$
replaced by $I_{\rm bias}=sI_c$. The time rate of change of $E$
reads
\begin{equation}\label{denergy}
   \frac{d}{dt}E = \left(\frac{\hbar}{2e}\right)^2\!
   C\dot\varphi\ddot\varphi + \frac{\hbar}{2e}\left[I_c
   \sin(\varphi)-I_{\rm bias}\right]\dot\varphi\, ,
\end{equation}
which, using the equation of motion (\ref{relaxation2}) satisfied 
by $\varphi_{\rm relax}(t)$, may be written as
\begin{equation}\label{denergyb}
   \frac{d}{dt}E = -\left(\frac{\hbar}{2e}\right)^2\!
   \frac{1}{R_{\vert\vert}}\dot\varphi^2 \, .
\end{equation}
Along the deterministic trajectory $\varphi_{\rm relax}(t)$ we may
look upon $E$ as a function of $\varphi$. Then
\begin{equation}\label{energyphase}
   \frac{dE}{d\varphi}=\frac{1}{\dot\varphi}\frac{dE}{dt}=-\left(\frac{\hbar}{2e}\right)^2\!
   \frac{1}{R_{\vert\vert}}\dot\varphi \, ,
\end{equation}
and from Eq.~(\ref{energy}) we have
\begin{equation}\label{dotphi}
\dot\varphi=\pm\frac{2e}{\hbar}\sqrt{\frac{2}{C}(E-U)}\, ,
\end{equation}
which combines with Eq.~(\ref{energyphase}) to yield
\begin{equation}\label{energyloss}
   \frac{dE}{d\varphi}=\pm\frac{\hbar}{2e}\frac{1}{R_{\vert\vert}}
   \sqrt{\frac{2}{C}\left(E-U\right)}\, ,
\end{equation}
where the sign is determined by the fact that $E$ decreases along
the trajectory.

The function $E(\varphi)$ can easily be determined by numerical
integration of Eq.~(\ref{energyloss}). One starts from
$\varphi=\varphi_{\rm top}$ with energy $E(\varphi_{\rm
top})=U(\varphi_{\rm top})$ and integrates towards smaller
$\varphi$ with the $+$ sign of Eq.~(\ref{energyloss}) until the
first turning point with $E(\varphi)=U(\varphi)$ is reached.
There, the integration continues towards larger values of
$\varphi$ with the $-$ sign of Eq.~(\ref{energyloss}) up to the
second turning point and so on, until the trajectory ends at
$E(\varphi_{\rm well})=U(\varphi_{\rm well})$.

By virtue of Eq.~(\ref{dotphi}) the formula (\ref{jrelax}) may be
written as
\begin{equation}\label{jrelaxa}
    J=-\frac{1}{3C}\left(\frac{2e}{\hbar}\right)^2\int_{-\infty}^{\infty} dt \,
    \dot\varphi\,
    (E-U)\, .
\end{equation}
Changing from an integration over time to one over phase, we get
\begin{equation}
\label{jfin}
J=-\frac{1}{3C}\left(\frac{2e}{\hbar}\right)^2\int_{\varphi_{\rm
top}}^{\varphi_{\rm well}} d\varphi\,
   (E-U)
   \, ,
\end{equation}
where the integration starts at $\varphi_{\rm top}$ and goes back
and forth between the turning points until it ends in
$\varphi_{\rm well}$. The determination of the effect of
non-Gaussian noise on the rate of escape is thus reduced to an
integration of the first order differential equation
(\ref{energyloss}) and the evaluation of the integral
(\ref{jfin}).
%%%%%%%%%%%%%%%%%%%%%%%%%%%%%%%%%%%%%%%
%%%%%%%%%%%%%%%%%%%%%%%%%%%%%%%%%%%%%%%
%%%%%      SECTION DIS           %%%%%%
%%%%%%%%%%%%%%%%%%%%%%%%%%%%%%%%%%%%%%%
%%%%%%%%%%%%%%%%%%%%%%%%%%%%%%%%%%%%%%%
\section{Discussion}
\label{sec_dis}

In this section we will give some concrete results in the
experimentally relevant range of parameters.
%%%%%%%%%%%%%%%%%%%%%%%%%%%%%%%%%%%%%%
%%%%%%%%%%%%%%%%%%%%%%%%%%%%%%%%%%%%%%
\subsection{Dimensionless Quantities}
It is convenient to formulate the theory in terms of dimensionless
quantities. Introducing the plasma frequency of the Josephson
junction at vanishing bias current
\begin{equation}
\label{plasmaf0}
   \omega_p=\sqrt{\frac{2e}{\hbar}\frac{I_c}{C}}\, ,
\end{equation}
the result (\ref{jfin}) may be written as
\begin{equation}\label{jdim}
   J=\omega_p^2j\, ,
\end{equation}
where
\begin{equation}\label{smallj}
   j=-\frac{1}{3}\int_{\varphi_{\rm
top}}^{\varphi_{\rm well}} d\varphi\,
   (e-u)
\end{equation}
is a dimensionless integral given in terms of the dimensionless
energy
\begin{equation}\label{smalle}
   e=\frac{2e}{\hbar}\frac{E}{I_c}=
   \frac{1}{2\omega_p^2}\dot\varphi^2+u
\end{equation}
and the dimensionless potential
\begin{equation}\label{smallu}
   u=\frac{2e}{\hbar}\frac{U}{I_c}=-\cos(\varphi)-s\varphi\, .
\end{equation}
From Eq.~(\ref{b3}), the correction $B_3$ to the exponential
factor of the rate may then be written as
\begin{equation}\label{b3fin}
   B_3=\left(\frac{\hbar}{2e}\right)^3\!
\frac{\omega_p^2}{\left(k_BT_{\rm eff}\right)^3}\, \mathcal{C}_3
j\, .
\end{equation}
To determine $j$ from Eq.~(\ref{smallj}), one needs to solve the
dimensionless form of Eq.~(\ref{energyloss}), which reads
\begin{equation}\label{smallephase}
\frac{de}{d\varphi}=\pm\gamma
   \sqrt{2(e-u)}\, ,
\end{equation}
where
\begin{equation}\label{gamma}
   \gamma=\frac{1}{R_{\vert\vert}C\omega_p}
\end{equation}
is the dimensionless damping coefficient, which coincides with the
inverse quality factor $Q=R_{\vert\vert}C\omega_p$ at vanishing
bias current.
%%%%%%%%%%%%%%%%%%%%%%%%%%%%%%%%%%%%%%
%%%%%%%%%%%%%%%%%%%%%%%%%%%%%%%%%%%%%%
\subsection{Strong Damping}

Let us first discuss the limit of strong damping $\gamma\gg 1$.
The Josephson junction noise detector cannot operate in this
limit, because after escape from the metastable well the phase
will be retrapped in the adjacent well of the tilted washboard
potential, so that only a short voltage pulse builds up.
Nevertheless, the behavior in this limit is instructive, since
explicit analytical results can be obtained. To solve
Eq.~(\ref{smallephase}), we make the ansatz
\begin{equation}\label{epsilon}
   e=u+\kappa
\end{equation}
and find
\begin{equation}\label{epsiloneq}
   \frac{d\kappa}{d\varphi}=-\frac{du}{d\varphi}\pm\gamma\sqrt{2\kappa}\,
   .
\end{equation}
This gives
\begin{equation}\label{epsilonsol}
   \sqrt{\kappa}=\pm\frac{1}{\sqrt{2}\,\gamma}\left(\frac{du}{d\varphi}
  + \frac{d\kappa}{d\varphi}\right)\, ,
\end{equation}
so that the dimensionless kinetic energy $\kappa$ is of order
$1/\gamma^2$ for large $\gamma$. The leading order solution
\begin{equation}\label{epsilonsolb}
   \kappa =
   \frac{1}{2\gamma^2}\left(\frac{du}{d\varphi}\right)^2
\end{equation}
satisfies the boundary conditions $e=u$, i.e. $\kappa=0$, for
$\varphi=\varphi_{\rm top}$ and $\varphi=\varphi_{\rm well}$.
Inserting Eq.~(\ref{epsilonsolb}) into Eq.~(\ref{smallj}), we
obtain
\begin{equation}\label{jstrong}
   j=-\frac{1}{6\gamma^2}\int_{\varphi_{\rm
top}}^{\varphi_{\rm well}} d\varphi\,
   \left(\frac{du}{d\varphi}\right)^2\, .
\end{equation}
In the overdamped limit, there are no turning points, but the
phase gradually slides down from $\varphi_{\rm top}$ to
$\varphi_{\rm well}$. Using Eqs.~(\ref{well}) and (\ref{smallu}),
Eq.~(\ref{jstrong}) is readily evaluated with the result
\begin{equation}\label{jstrongex}
   j=\frac{\left(1+2s^2\right)\arccos(s)-3s\,\sqrt{1-s^2}}{6\gamma^2}\,
   .
\end{equation}
Now, the observed escape events occur typically for values of the
bias current $I_{\rm bias}$ close to the critical current $I_c$.
Then, $1-s\ll 1$ and Eq.~(\ref{jstrongex}) can be expanded to
yield
\begin{equation}\label{jstrongapp}
   j=\frac{8\sqrt{2}}{45}\left(1-s\right)^{5/2}\frac{1}{\gamma^2}\, .
\end{equation}
This latter formula is in accordance with the result by Sukhorukov
and Jordan\cite{Sukhorukov} in this limit.
%%%%%%%%%%%%%%%%%%%%%%%%%%%%%%%%%%%%%%
%%%%%%%%%%%%%%%%%%%%%%%%%%%%%%%%%%%%%%
\subsection{Very Weak Damping}
Next we consider the case of a very weakly damped Josephson
junction, i.e., $\gamma\ll 1$. Then the trajectory $\varphi(t)$
oscillates back and forth in the potential well and looses energy
only very gradually. Let us consider a segment of the trajectory
starting at a turning point $\varphi_+$ on the barrier side of the
potential, oscillating through the potential well to a turning
point $\varphi_-$ on the opposite side, and traversing the
potential well again to a turning point $\varphi_+^{\prime}$. From
Eq.~(\ref{smallephase}) we find for the energy along this path
segment
\begin{eqnarray}\label{ephi}
\nonumber
   e(\varphi) = e(\varphi_+)
   &+& \gamma \int_{\varphi_+}^{\varphi_-}  d\varphi
\sqrt{2\left(e-u\right)}\\ \nonumber \\
&\pm& \gamma \int_{\varphi_-}^{\varphi}  d\varphi
\sqrt{2\left(e-u\right)}\, ,
\end{eqnarray}
where the $+$ sign holds for the oscillation form $\varphi_+$ to
$\varphi_-$, and the $-$ sign on the way back from $\varphi_-$ to
$\varphi_+^{\prime}$. For $\gamma\ll 1$, this gives
\begin{eqnarray}\label{ephip}
\nonumber
   e(\varphi) = e_+
   &+& \gamma \int_{\varphi_+}^{\varphi_-}  d\varphi
\sqrt{2\left(e_+-u\right)}\\ \nonumber \\
&\pm& \gamma \int_{\varphi_-}^{\varphi}  d\varphi
\sqrt{2\left(e_+-u\right)} +\mathcal{O}(\gamma^2)\, ,\quad
\end{eqnarray}
where $e_+=e(\varphi_+)=u(\varphi_+)$. This result can now be 
inserted into Eq.~(\ref{smallj}), to find for a segment of the 
$\varphi$-integral form $\varphi_+$ over $\varphi_-$ to 
$\varphi_+^{\prime}$
\begin{eqnarray}
\label{deltaj}
 \Delta j &=& -\frac{1}{3}\left\{ \int_{\varphi_+}^{\varphi_-}d\varphi\,
   (e-u) +\int_{\varphi_-}^{\varphi_+^{\prime}} d\varphi\,
   (e-u)\right\} \\
  &&\nonumber  \\
  &=&\frac{2}{3}\gamma\int_{\varphi_-}^{\varphi_+}d\varphi
  \int_{\varphi_-}^{\varphi}d\varphi^{\prime}
  \sqrt{2\left[e_+-u(\varphi^{\prime})\right]}
  +\mathcal{O}(\gamma^2)\,
,\nonumber
\end{eqnarray}
where we have taken into account that the difference between
$\varphi_+$ and $\varphi_+^{\prime}$ is of order $\gamma$.

On the other hand, Eq.~(\ref{ephip}) gives for the change $\Delta
e$ of the energy during one oscillation period
\begin{equation}\label{eloss}
   \Delta e = -2\gamma\int_{\varphi_-}^{\varphi_+}d\varphi
   \sqrt{2\left[e_+-u(\varphi)\right]}
   +\mathcal{O}(\gamma^2)\, .
\end{equation}
Eqs.~(\ref{deltaj}) and (\ref{eloss}) combine to yield
\begin{equation}\label{djde}
   \frac{\Delta j}{\Delta e}=-f(e) +\mathcal{O}(\gamma)\,,
\end{equation}
where
\begin{equation}\label{fvone}
   f(e)= \frac{1}{3}\frac{\int_{\varphi_-}^{\varphi_+}d\varphi
  \left(\varphi_+-\varphi\right)
  \sqrt{e-u(\varphi)}}{\int_{\varphi_-}^{\varphi_+}d\varphi
   \sqrt{e-u(\varphi)}}\, .
\end{equation}
Dividing the integral (\ref{smallj}) into segments of the form
(\ref{deltaj}), we can transform the integral over $\varphi$ into
an integral over energy. Using Eq.~(\ref{djde}), we then obtain
\begin{equation}\label{jweak}
   j=\int_{u(\varphi_{\rm well})}^{u(\varphi_{\rm top})} de f(e)\,
   .
\end{equation}

Let us again study specifically the experimentally important range
$1-s\ll 1$. Then, the relevant range of $\varphi$ values lies in
the vicinity of $\frac{\pi}{2}$. Putting
\begin{equation}\label{psi}
   \varphi=\frac{\pi}{2}+\sqrt{2(1-s)}\,\psi\, ,
\end{equation}
we find for the potential (\ref{smallu})
\begin{equation}\label{poteps}
   u=-\frac{\pi}{2}s+\sqrt{2}(1-s)^{3/2}\varsigma\, ,
\end{equation}
where
\begin{equation}\label{sigma}
   \varsigma=\psi-\frac{1}{3}\psi^3 \, .
\end{equation}
With the scaled dimensionless energy
\begin{equation}\label{esig}
   e=-\frac{\pi}{2}s+\sqrt{2}(1-s)^{3/2}\epsilon\, ,
\end{equation}
the result (\ref{jweak}) with (\ref{fvone}) can be transformed to
read
\begin{eqnarray}\label{jweaks}
\nonumber
   j&=&\frac{2}{3}(1-s)^2
   \\ \nonumber \\
   &\times& \int_{-\frac{2}{3}}^{\frac{2}{3}}d\epsilon\,
   \frac{\int_{\psi_-}^{\psi_+}d\psi \left(\psi_+-\psi\right)
   \sqrt{\epsilon-\varsigma(\psi)}}{\int_{\psi_-}^{\psi_+}d\psi
   \sqrt{\epsilon-\varsigma(\psi)}}\, ,
\end{eqnarray}
where $\psi_-$ and $\psi_+$ are the negative and smallest positive
roots of $\varsigma(\psi)=\psi-\frac{1}{3}\psi^3=\epsilon$,
respectively. The remaining integral is just a numerical factor
independent of $s$, and a numerical evaluation gives
\begin{equation}\label{jweakse}
   j=a(1-s)^2 ,\, \hbox{with}\quad a=0.79\ldots\, .
\end{equation}
This result is in accordance with the findings by Sukhorukov and
Jordan\cite{Sukhorukov} in the limit of vanishing damping.
%%%%%%%%%%%%%%%%%%%%%%%%%%%%%%%%%%%%%%
%%%%%%%%%%%%%%%%%%%%%%%%%%%%%%%%%%%%%%
\subsection{Intermediate Damping}

\begin{figure}
\vspace{0.3cm}
\begin{center}
\includegraphics[width=0.9\columnwidth]{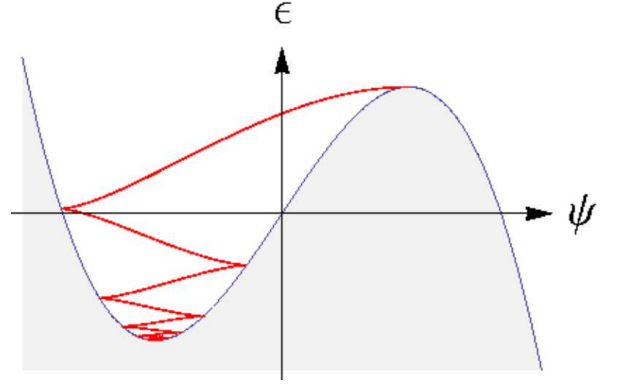}
\caption{\label{trajectory}
The scaled dimensionless energy $\epsilon$ is shown as a function
of $\psi$ for $\gamma=0.25$. The energy decreases as the trajectory
moves back and forth in the potential $\varsigma(\psi)$ depicted
as a grey line.}
\end{center}
\end{figure}

In experiments typical values of the dimensionless damping
coefficient $\gamma$ are small but nonvanishing. The factor $j$ in
formula (\ref{b3fin}) for $B_3$ must then be determined from
Eq.~(\ref{smallj}) using the solution of the differential equation
(\ref{smallephase}). While a numerical evaluation is
straightforward for arbitrary values of $s$, we shall focus on the
experimentally relevant range $1-s\ll 1$. In terms of the scaled
quantities introduced in Eqs.~(\ref{psi})\,--\,(\ref{esig}),
Eq.~(\ref{smallephase}) reads
\begin{equation}\label{scaleddiff}
   \frac{d\epsilon}{d\psi}=\pm\tilde\gamma\, \sqrt{2(\epsilon-\varsigma)}\,
   ,
\end{equation}
where
\begin{equation}\label{tildegamma}
   \tilde\gamma=\left(\frac{2}{1-s}\right)^{\frac{1}{4}}\!\gamma\,
   .
\end{equation}
This differential equation has to be solved with initial condition
$\epsilon(1)=\varsigma(1)=\frac{2}{3}$, and integrated with the
proper sign back and forth between the turning points until the
integration ends at $\epsilon(-1)=\varsigma(-1)=-\frac{2}{3}$. A
typical solution is depicted in Fig.~\ref{trajectory}.
%%%%%%%%%%%%%%%%%%%%%%%%%%%
%%%%%%%%%%%%%%%%%%%%%%%%%%%
%%%%%%               %%%%%%
%%%%%%   Figure 3    %%%%%%
%%%%%%               %%%%%%
%%%%%%%%%%%%%%%%%%%%%%%%%%%
%%%%%%%%%%%%%%%%%%%%%%%%%%%
In scaled units Eq.~(\ref{smallj}) takes the form
\begin{equation}\label{scalej}
   j=-\frac{2}{3}(1-s)^2\int_{1}^{-1}d\psi
   (\epsilon-\varsigma)\, ,
\end{equation}
where the integral follows the $\psi$-path back and forth between
the turning points. Since the differential equation
(\ref{scaleddiff}) depends on $s$ and $\gamma$ only in the
combination $\tilde\gamma$, we put
\begin{equation}\label{smalljw}
   j=\frac{2}{3}(1-s)^2W(\tilde\gamma)\, ,
\end{equation}
where\begin{equation}\label{wgamma}
   W(\tilde\gamma)=-\int_{1}^{-1}d\psi
   (\epsilon-\varsigma)\, .
\end{equation}
The function $W(\tilde\gamma)$ determines the correction $B_3$ of
the exponential factor of the rate for arbitrary damping strength
in the range of bias currents close to the critical current.

\begin{figure}
\vspace{0.3cm}
\begin{center}
\includegraphics[scale=0.3]{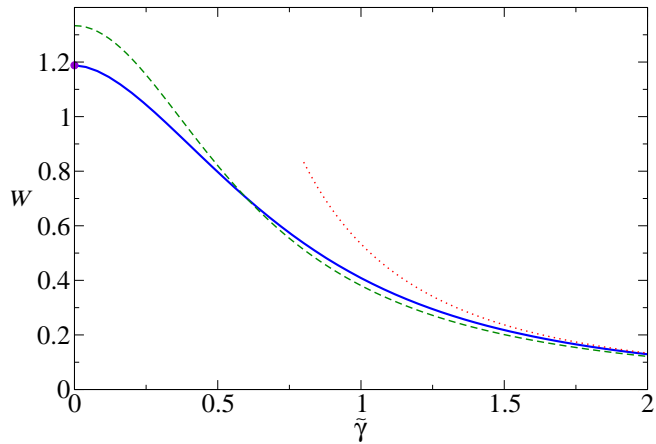}
\caption{\label{plot} $W$ is depicted as a function of 
$\tilde\gamma$ (straight line). Also shown are the results of 
Ref.~\onlinecite{Sukhorukov} for vanishing damping, 
Eq.~(\ref{wo}), (dot), and in the strong damping limit, 
Eq.~(\ref{wstrong}), (dotted line). The approximate result of 
Ref.~\onlinecite{Ankerhold2} is depicted as a dashed line.}
\end{center}
\end{figure}

From Eq.~(\ref{jweakse}), we obtain
\begin{equation}\label{wo}
   W(0)= 1.188\ldots ,
\end{equation}
while Eq.~(\ref{jstrongapp}) gives for $\tilde\gamma\gg 1$
\begin{equation}\label{wstrong}
   W(\tilde\gamma)\approx \frac{8}{15}\,\frac{1}{\tilde\gamma^2}\, ,
\end{equation}
where we have made use of Eq.~(\ref{tildegamma}). In between 
these limiting results, the function needs to be determined 
numerically. A list of data points is provided in 
Table~\ref{wtable}, and the function $W(\tilde\gamma)$ is 
depicted in Fig.~\ref{plot} together with the findings of 
previous works.\cite{Ankerhold2,Sukhorukov} This should 
facilitate the comparison with experimental results.
%%%%%%%%%%%%%%%%%%%%%%%%%%%
%%%%%%%%%%%%%%%%%%%%%%%%%%%
%%%%%%               %%%%%%
%%%%%%     Table     %%%%%%
%%%%%%               %%%%%%
%%%%%%%%%%%%%%%%%%%%%%%%%%%
%%%%%%%%%%%%%%%%%%%%%%%%%%%
\begin{table}
 \centering
 \begin{ruledtabular}
 \begin{tabular}{c|c|c|c|c|c|c|c|c}
   $\tilde\gamma$ & 0 & 0.025 & 0.05 & 0.075 & 0.1 & 0.125 & 0.15 & 0.175  \\
   \hline
   $W$ & 1.188 &   1.185 & 1.179 & 1.169 & 1.157 & 1.142 & 1.125 & 1.107   \\
 \end{tabular}

\vspace{0.3cm}
 \begin{tabular}{c|c|c|c|c|c|c|c|c}
   $\tilde\gamma$& 0.2 & 0.225 & 0.25 & 0.5 & 0.75 & 1.0 & 1.5 & 2.0 \\
   \hline
   $W$ & 1.087 & 1.066 & 1.043 &  0.797 & 0.574 & 0.409 & 0.218 & 0.129 \\
 \end{tabular}
  \end{ruledtabular}
 \caption{Some numerical values for $W$ as a function of $\tilde\gamma$.}\label{wtable}
\end{table}
%%%%%%%%%%%%%%%%%%%%%%%%%%%
%%%%%%%%%%%%%%%%%%%%%%%%%%%
%%%%%%               %%%%%%
%%%%%%   Figure 4    %%%%%%
%%%%%%               %%%%%%
%%%%%%%%%%%%%%%%%%%%%%%%%%%
%%%%%%%%%%%%%%%%%%%%%%%%%%%

%%%%%%%%%%%%%%%%%%%%%%%%%%%%%%%%%%%%%%
%%%%%%%%%%%%%%%%%%%%%%%%%%%%%%%%%%%%%%
%%%%%%%%%%%%%%%%%%%%%%%%%%%%%%%%%%%%%%
\subsection{Conclusions}

We have presented a theory for a Josephson junction detecting
non-Gaussian fluctuations by means of the noise driven escape out
of the zero-voltage state of the junction. It has been assumed
that the device is operated in a regime where the barrier of the
washboard potential is overcome by activated processes. This is
always the case if the temperature is not too low and/or the
junction capacitance is not too small. The study was based on the
theory of irreversible processes and fluctuations developed by
Onsager and Machlup\cite{Onsager} and Grabert, Graham, and
Green.\cite{GG,GGG} An extension of the method to account for
non-Gaussian fluctuations was outlined.\cite{else} In this 
approach the random motion of the system is described in terms of 
the state variables and the conjugate forces. The force $\lambda$ 
conjugate to the electric charge $Q$, which appears naturally in 
this approach, plays a role similar to the counting field 
introduced in the more recent approaches to determine the full 
counting statistics of electronic devices.\cite{reviews}

A nonlinear noise generating element in series with the Josephson
detector modifies the rate of escape out of the zero-voltage
state. The main effect comes from the second noise cumulant
$C_{2,N}$. However, this Gaussian part of the noise is detected by
the Josephson junction in the same way as Johnson-Nyquist noise.
Therefore, as was shown explicitly, the second noise cumulant can
be described in terms of an effective temperature $T_{\rm eff}$.
Deviations from the accordingly modified Arrhenius law are thus
due to higher order noise cumulants. The fluctuations causing the
escape from the metastable well lead to fluctuations of the
voltage $V_J$ across the Josephson junction. It has been shown
that these voltage fluctuations are small compared to $k_BT_{\rm
eff}/e$, which implies that the dimensionless random force
$e\lambda/k_B$ causing these fluctuations is always small compared
to 1. Since the $n^{\rm th}$ order noise cumulant gives rise to
terms of order $(e\lambda/k_B)^n$, deviations from the modified
Arrhenius law essentially only arise from the third noise cumulant
$C_{3,N}$, and these corrections are typically small. However, the
third cumulant is odd under time reversal and the sign of the
effect depends on the direction of the bias current. Comparing
rates for pulses tilting the potential to the right and the left,
respectively, the correction $B_3$ can be
extracted.\cite{Pekola,Pothier} A Josephson junction threshold
detector operating in the regime of noise activated escape thus 
can measure the third cumulant, the skewness of the noise, only. 
Another effect of the fluctuations of the voltage $V_J$ is a 
feedback of the Josephson detector on the noise generating device 
as described by the effective third noise cumulant 
$\mathcal{C}_3$ defined in Eq.~(\ref{c3a}).

The modification of the rate exponent due to the skewness of the
noise has been determined for arbitrary damping strength of the
Josephson junction detector. Thereby, the theory developed goes
considerably beyond the results of previous
works,\cite{Ankerhold2,Sukhorukov} that were restricted to
limiting values of the damping strength or based on
approximations. Explicit results where given for the case when
the bias current is close to the critical current, which implies
that the relevant part of the washboard potential can be
described by a cubic potential. The effect of the skewness of the
noise on the rate is, however, larger for smaller values of the
bias current. Experimentally, the range of relevant bias currents
can be influenced by the form of the applied current pulses. The 
theory presented here can readily also be evaluated for the exact 
form of the washboard potential allowing for results for any 
value of the bias current and all damping strengths.

To be explicit, we have presented the theory using the example of
a normal state tunnel junction as noise generating device.
However, the theory readily also applies to other noise generating
elements, provided the correlation time of the noise is much
smaller than the period of plasma oscillations of the detector. 
Finally, in this article, only the exponential factor of the rate 
has been determined. The corrections due to the skewness of the 
noise were found to be rather small, and they need sophisticated 
experimental techniques to be detected reliably. Corrections to 
the pre-exponential factor of the same order of magnitude are 
entirely negligible, so that safely the prefactor of the standard 
Gaussian noise theory can be employed.

%%%%%%%%%%%%%%%%%%%%%%%%%%%%%%%%%%%%%%
%%%%%%%%%%%%%%%%%%%%%%%%%%%%%%%%%%%%%%
%%%%%%%%%%%%%%%%%%%%%%%%%%%%%%%%%%%%%%
\begin{acknowledgments}
This work was carried out in the summer of 2007 during a
sabbatical visit to CEA-Saclay. The warm hospitality of the
Quantronics group and the enlightening discussions with the group
members, in particular with  D. Est\`eve and H. Pothier, are
gratefully acknowledged.  The author also wishes to thank J.
Ankerhold and J. Pekola for a number of interesting discussions, 
as well as D. Bercioux for discussions and providing the figures. 
Financial support was allocated by the European NanoSci-ERA 
Programme.
\end{acknowledgments}

\appendix
%%%%%%%%%%%%%%%%%%%%%%%%%%%%%%%%%%%%%%%%%%%%%%%%%
%%%%%%%%%%%%%%%%%%%%%%%%%%%%%%%%%%%%%%%%%%%%%%%%%
%%%%%%%%%%%%%%%%%%%%%%%%%%%%%%%%%%%%%%%%%%%%%%%%%
%%%%%%       A P P E N D I X    A         %%%%%%%
%%%%%%%%%%%%%%%%%%%%%%%%%%%%%%%%%%%%%%%%%%%%%%%%%
%%%%%%%%%%%%%%%%%%%%%%%%%%%%%%%%%%%%%%%%%%%%%%%%%
%%%%%%%%%%%%%%%%%%%%%%%%%%%%%%%%%%%%%%%%%%%%%%%%%

\section{Validity of Nearly Gaussian Approximation}
\label{appa}

In this appendix we investigate the range of validity of the
nearly Gaussian approximation used in Sec.~\ref{sec_NG}. Since the
leading order term $\varphi_2(t)$ of the most probable escape path
is the time reversed relaxation path $\varphi_{\rm relax}(t)$, the
order of magnitude of the phase velocity $\dot \varphi$ during
escape coincides with that during relaxation.

Let us first consider the case of weak damping. The trajectory
$\varphi_{\rm relax}(t)$ starts with vanishing phase velocity at
the barrier top. The largest kinetic energy
$\frac{1}{2}\left(\hbar/2e\right)^2C\,\dot \varphi^2$ arises when
the potential minimum $\varphi_{\rm well}$ is reached for the
first time. For weak damping the kinetic energy then almost
equals the potential energy difference $\Delta U$. Accordingly,
the voltage $V_J=\left(\hbar/2e\right)\dot \varphi$ satisfies
\begin{equation}\label{vjot}
   V_J \le \sqrt{\frac{2\Delta U}{C}}\, .
\end{equation}
As damping increases the phase velocity and, accordingly, the
maximal voltage across the Josephson junction decreases, so that
$V_J$ will never exceed the estimate (\ref{vjot}) in the entire
range of parameters.

The plasma frequency of the Josephson junction at finite bias
current
\begin{equation}\label{plasmaf}
   \omega_p(s)=\omega_p\sqrt{\sin(\delta)}=\sqrt{\frac{2e}{\hbar}\frac{I_c}{C}\sin(\delta)}
\end{equation}
is the frequency of small undamped oscillations about the minimum
$\varphi_{\rm well}$ of the potential (\ref{potential}). For
$\delta \ll \frac{\pi}{2}$, which is the case for $1-s\ll 1$,
Eqs.~(\ref{potential}) - (\ref{delta}) yield for the barrier
height (\ref{barrierheight}) of the potential
\begin{equation}\label{barrierheight2}
   \Delta U\approx \frac{\hbar I_c}{3e}\delta^3\, .
\end{equation}
This can be combined with Eq.~(\ref{plasmaf}) to give
\begin{equation}
\label{estimate}
   \hbar\omega_p(s)\approx \sqrt{
   \frac{2e\hbar\, I_c\,\delta}{C}} \approx
   \frac{e}{\delta}\sqrt{ \frac{6\Delta U}{C}}\, .
\end{equation}
The bound (\ref{vjot}) for the size of the fluctuations of $V_J$
may thus be written as
\begin{equation}\label{vjot2}
   eV_J\le \frac{\delta}{\sqrt{3}}\,\hbar\omega_p(s)\, .
\end{equation}

In the region of thermally activated escape\cite{remark5} one has
$\hbar\omega_p(s)\ll k_BT_{\rm eff}$. In view of Eq.~(\ref{vjot2})
this implies
\begin{equation}\label{para1}
   \frac{eV_J}{k_BT_{\rm eff}}\ll\frac{\delta}{\sqrt{3}}\ll 1\, ,
\end{equation}
so that $eV_J/k_BT_{\rm eff}$ is a small dimensionless parameter
along the most probable escape path.

Now, the leading order contribution $\lambda_2$ to the force
$\lambda$ causing the escape is determined by Eq.~(\ref{phi2}),
entailing the estimate
\begin{equation}\label{lamdaest}
   \lambda \approx \frac{\hbar}{eT_{\rm eff}}\dot\varphi \approx
   \frac{2V_J}{T_{\rm eff}}\, ,
\end{equation}
which combines with the inequality (\ref{para1}) to give
\begin{equation}\label{lambdaest2}
   \frac{e\lambda}{k_B}\ll 1\,.
\end{equation}
This shows that an expansion of the Hamiltonian in terms of
$\lambda$, as done in Eq.~(\ref{hamiltonianng}), is indeed
justified. The terms of third order in $\lambda$ are then small,
so that $\varphi_3$ and $\lambda_3$ describe in fact small
corrections to $\varphi_2$ and $\lambda_2$, respectively.

Because of the weak effects of non-Gaussian statistics, the
correction $B_3$ to the exponent of the rate is also small. From
Eqs.~(\ref{b2}) and (\ref{b3fin}), we find
\begin{equation}\label{bratio}
   \frac{B_3}{B_2} =\frac{\hbar}{(2e)^3}\left(\frac{\hbar\omega_p}{k_BT_{\rm
   eff}}\right)^2 \frac{\mathcal{C}_3}{\Delta U} \, j\, .
\end{equation}
For $\delta \ll \frac{\pi}{2}$ we can insert Eqs.~(\ref{smalljw})
and (\ref{barrierheight2}). Using Eq.~(\ref{delta}), we then find
\begin{equation}\label{bratio2}
   \frac{B_3}{B_2} \approx \frac{1}{16}\left(\frac{\hbar\omega_p}{k_BT_{\rm
   eff}}\right)^2 \frac{\mathcal{C}_3}{e^2I_c} \,  W\,\delta\, .
\end{equation}
Hence, the effect of the skewness of the noise vanishes
proportional to $(1-s)^{1/2}$ as the bias current approaches
$I_c$. The ratio $B_3/B_2$ can be seen as a product of three
factors
\begin{equation}\label{bratio3}
   \frac{B_3}{B_2} \approx \frac{1}{16}\left(\frac{\hbar\omega_p(s)}{k_BT_{\rm
   eff}}\right)^2 \!\times \frac{\mathcal{C}_3}{e^2I_c} \times W\, ,
\end{equation}
where we have made use of Eq.~(\ref{plasmaf}). Now, in the regime
of activated decay the first factor
$(1/16)(\hbar\omega_p(s)/k_BT_{\rm eff})^2$ is very small, while
the last factor $W$ is of order 1 for weak to moderate damping.
Hence, one needs a large factor $\mathcal{C}_3/e^2I_c$ to get
observable effects from the skewness of the noise. Since
$C_{3,N}$ is proportional to $V_N$, this means large $V_N$, in
particular, $eV_N\gg k_BT$, so that the estimate (\ref{c3approx})
for $\mathcal{C}_3$ applies.  To minimize the reduction of
$C_{3,N}$ via the feedback effects described by
Eq.~(\ref{c3approx}), one needs to choose a bias resistor $R_B$
well below $R_N$. Then the factor
\begin{equation}\label{factor}
 \frac{\mathcal{C}_3}{e^2I_c} \approx \frac{C_{3,N}}{e^2I_c} = \frac{V_N}{R_NI_c}\, .
\end{equation}
This means that the current $V_N/R_N$ should be large compared to
$I_c$  and thus needs to be largely compensated by a current
$V_B/R_B$ in the opposite direction to keep the junction biasing
current (\ref{bias}) below $I_c$. Experimentally, this
compensation problem is addressed by employing more sophisticated
set-ups.\cite{Pekola,Pothier}

%%%%%%%%%%%%%%%%%%%%%%%%%%%%%%%%%%%%%
%%%%%%%%%%%%%%%%%%%%%%%%%%%%%%%%%%%%%
%%%%%%%%%%%%%%%%%%%%%%%%%%%%%%%%%%%%%
%%%%%%%%%%%%%%%%%%%%%%%%%%%%%%%%%%%%%
%%%%%%%%%%%%%%%%%%%%%%%%%%%%%%%%%%%%%
\section{Evaluation of Action of Escape Path}
\label{appb} In this Appendix we evaluate the expressions
(\ref{a2}) and (\ref{a3}) for the action of the escape path in the
nearly Gaussian approximation. Inserting the result (\ref{phi2})
for $\lambda_2$, one obtains from (\ref{a2})
\begin{eqnarray}\label{a2b}
   \nonumber
   A_2&=&\frac{2}{T_{\rm eff}}\int_{-\infty}^{\infty} \!dt\left[
   \left(\frac{\hbar}{2e}\right)^2\!C\dot\varphi_2\ddot\varphi_2
    - \frac{\hbar}{2e}I_c \cos(\varphi_2)\varphi_2\dot\varphi_2\right]\\
    \nonumber\\
    \nonumber
    &=&\frac{2}{T_{\rm eff}}\int_{-\infty}^{\infty} dt\bigg\{
   \frac{\partial}{\partial t}\frac{1}{2}\left(\frac{\hbar}{2e}\right)^2\!C\dot\varphi_2^2
   \\
   \nonumber\\
    &&- \frac{\partial}{\partial t}
    \frac{\hbar}{2e}I_c
    \left[\cos(\varphi_2)+\varphi_2\sin(\varphi_2)\right]\bigg\} .
\end{eqnarray}
Now, $\dot\varphi_2$ vanishes at the integration boundaries and
$-(\hbar/2e)I_c
\left[\cos(\varphi_2)+\varphi_2\sin(\varphi_2)\right]$ coincides
there with $U(\varphi_{\rm top})$ and $U(\varphi_{\rm well})$,
respectively, since $\sin(\varphi_{\rm well})=\sin(\varphi_{\rm
top})=s$. Accordingly, Eq.~(\ref{a2b}) yields
\begin{equation}\label{a2c}
   A_2=\frac{2\Delta U}{T_{\rm eff}}\, ,
\end{equation}
which gives the exponential factor (\ref{b2}) of the escape rate.

After expressing $\lambda_2$ in terms of $\varphi_2$ and putting
\begin{equation}\label{lambdatilde}
   \lambda_3=\frac{\hbar}{eT_{\rm eff}} \Lambda_3\, ,
\end{equation}
we obtain from Eq.~(\ref{a3}) for the leading order non-Gaussian
part of the action
\begin{eqnarray}
\label{a3b}
   A_3&=&\frac{1}{T_{\rm eff}}\int_{-\infty}^{\infty} dt\bigg\{
    \frac{1}{2}\left(\frac{\hbar}{e}\right)^2\!C\left(\dot\varphi_2\ddot\varphi_3
    +\ddot\varphi_2\Lambda_3 \right)\\
    &-& \frac{\hbar}{e}I_c\left[ \cos(\varphi_2)\left(\dot\varphi_2\varphi_3+\varphi_2\Lambda_3\right)
    -\sin(\varphi_2)\varphi_2\dot\varphi_2\varphi_3\right]
    \bigg\}\, .\nonumber
\end{eqnarray}
The integral in the first line gives after partial integration
\begin{equation}\label{a3c}
  A_{3,\rm part\, I}=\frac{1}{T_{\rm eff}}\int_{-\infty}^{\infty} dt
    \frac{1}{2}\left(\frac{\hbar}{e}\right)^2\!C\left(\dot\varphi_2\ddot\varphi_3
    +\varphi_2\ddot\Lambda_3 \right)\, .
\end{equation}
In this expression we can eliminate the second order derivatives
$\ddot\varphi_3$ and $\ddot\Lambda_3$ by means of the equations of
motion (\ref{eqmphi3}) and (\ref{eqmlambda3}). Taking the
definition (\ref{lambdatilde}) into account, we get
\begin{eqnarray}\label{a3d}
\nonumber
  A_{3,\rm part\, I}&=&\frac{\hbar}{eT_{\rm eff}}\int_{-\infty}^{\infty} dt\bigg\{
  \dot\varphi_2\bigg[-\frac{\hbar}{2e}\frac{1}{R_{\vert\vert}}\dot\varphi_3\\
  \nonumber\\
    &&-I_c\cos(\varphi_2)\varphi_3+\frac{\hbar}{eR_{\vert\vert}}\Lambda_3+I_3 \bigg]
   \nonumber\\
   \nonumber\\
    &&+\varphi_2\bigg[\frac{\hbar}{2e}\frac{1}{R_{\vert\vert}}\dot\Lambda_3
   -I_c\cos(\varphi_2)\Lambda_3
   \nonumber\\
   \nonumber\\
   &&+\dot \varphi_2\left(I_c\sin(\varphi_2)\varphi_3+I_3^{\prime}\right)\bigg] \bigg\}\, .
\end{eqnarray}
This result can now be inserted into (\ref{a3b}). After a partial
integration of the $\varphi_2\dot\Lambda_3$ term and a further
partial integration along the lines $
I_c[\sin(\varphi_2)\varphi_2\dot\varphi_2
-\cos(\varphi_2)\dot\varphi_2]\varphi_3 = I_c[-(\partial/\partial
t)\cos(\varphi_2)\varphi_2]\varphi_3 \to
I_c\cos(\varphi_2)\varphi_2\dot\varphi_3$, one obtains
\begin{eqnarray}\label{a3e}
  A_3&=&\frac{\hbar}{eT_{\rm eff}}\int_{-\infty}^{\infty} dt\bigg\{
  \dot\varphi_2\left(I_3+\varphi_2I_3^{\prime}\right)
  \\
  \nonumber\\
   &&+\left(\frac{\hbar}{2e}\frac{1}{R_{\vert\vert}}
   \dot\varphi_2- 2I_c\cos(\varphi_2)\varphi_2  \right)
   \left(\Lambda_3-\dot\varphi_3 \right)\bigg\}
    \nonumber\, .
\end{eqnarray}
From Eqs.~(\ref{i30}) and (\ref{i3strich0}), we see that
\begin{eqnarray}\label{i3comb}
  \dot\varphi_2\left( I_3+\varphi_2I_3^{\prime}\right)
   &=&  \frac{1}{2} \left(\frac{\hbar}{2e}\right)^2 \frac{1}{\left(k_BT_{\rm
   eff}\right)^2} \, C_{3,N}\, \dot\varphi_2^3\\
   \nonumber\\
&-& \frac{1}{k_BT_{\rm eff}}
\left(\frac{\hbar}{2e}\right)^2\!
\frac{\partial C_{2,N}}{\partial V_N} \left(
    \dot\varphi_2^3
  - \varphi_2 \dot\varphi_2\ddot\varphi_2\right)\, .
  \nonumber
\end{eqnarray}
Now, under the integral $\varphi_2 \dot\varphi_2\ddot\varphi_2 =
\varphi_2 (\partial/\partial t)\frac{1}{2}\dot\varphi_2^2 \to -
\frac{1}{2}\dot\varphi_2^3$, so that $\dot\varphi_2\left(
I_3+\varphi_2I_3^{\prime}\right)$ can be replaced
by\begin{equation}\label{i3combib}
   \dot\varphi_2\left( I_3+\varphi_2I_3^{\prime}\right) \to
\frac{1}{2}  \left(\frac{\hbar}{2e}\right)^2
\frac{1}{\left(k_BT_{\rm eff}\right)^2}\, \mathcal{C}_3\,
\dot\varphi_2^3
\end{equation}
where
\begin{equation}\label{c3}
  \mathcal{C}_3= C_{3,N}- 3k_BT_{\rm eff}\frac{\partial C_{2,N}}{\partial
  V_N}\, .
\end{equation}
Since the action (\ref{a3e}) depends on $\Lambda_3-\dot\varphi_3$
only, is is natural to make the ansatz
\begin{equation}\label{ansatzlambda3}
   \Lambda_3=\dot\varphi_3+\Lambda_3^{\prime}\, .
\end{equation}
From Eq.~(\ref{lambdatilde}) and the equations of motion
(\ref{eqmphi3}) and (\ref{eqmlambda3}) one then finds
\begin{equation}\label{eqmlamda3strich}
   \frac{\hbar}{2e}C\ddot\Lambda_3^{\prime}
   +\frac{\hbar}{2e}\frac{1}{R_{\vert\vert}}\dot\Lambda_3^{\prime}
   +I_c\cos(\varphi_2)\Lambda_3^{\prime}=\dot\varphi_2I_3^{\prime}-\dot I_3\, .
\end{equation}
Using Eqs.~(\ref{i30}) and (\ref{i3strich0}), the right hand side
may be written as
\begin{equation}\label{inhom}
   \dot\varphi_2I_3^{\prime}-\dot I_3=
   -\left(\frac{\hbar}{2e}\right)^2\frac{1}{\left(k_BT_{\rm eff}\right)^2}\, \mathcal{C}_3\,
   \dot\varphi_2\ddot\varphi_2\, ,
\end{equation}
where again the cumulants appear only in the combination
(\ref{c3}).

We can now employ the evolution equation (\ref{eqmlamda3strich})
to express the term proportional to $I_c$ in the action
(\ref{a3e}) in favor of terms with a purely polynomial dependence
on $\varphi_2$. Using also Eqs.~(\ref{i3combib}),
(\ref{ansatzlambda3}), and (\ref{inhom}), we find
\begin{eqnarray}\label{a3f}
\nonumber
  A_3&=&\frac{\hbar}{eT_{\rm eff}}\int_{-\infty}^{\infty} dt\bigg\{
  \frac{1}{2}  \left(\frac{\hbar}{2e}\right)^2
\frac{1}{\left(k_BT_{\rm eff}\right)^2} \, \mathcal{C}_3\,
\dot\varphi_2^3
  \\
  \nonumber\\
  \nonumber
   &&+\frac{\hbar}{2e}\frac{1}{R_{\vert\vert}}
   \dot\varphi_2\Lambda_3^{\prime}+ 2\varphi_2\bigg[ \frac{\hbar}{2e}C\ddot\Lambda_3^{\prime}
   +\frac{\hbar}{2e}\frac{1}{R_{\vert\vert}}\dot\Lambda_3^{\prime}
\\
  \nonumber\\
    &&+\left(\frac{\hbar}{2e}\right)^2\left(\frac{1}{k_BT_{\rm
   eff}}\right)^2\mathcal{C}_3\,
   \dot\varphi_2\ddot\varphi_2\bigg]\bigg\}
    \, .
\end{eqnarray}
After partial integrations along the lines
$\varphi_2\ddot\Lambda_3^{\prime} \to
\ddot\varphi_2\Lambda_3^{\prime}$, $
\varphi_2\dot\Lambda_3^{\prime} \to
-\dot\varphi_2\Lambda_3^{\prime}$, and
$\varphi_2\dot\varphi_2\ddot\varphi_2 = \varphi_2(\partial/
\partial t) \frac{1}{2}\dot\varphi_2^2\to -\frac{1}{2}
\dot\varphi_2^3$, this simplifies to read
\begin{eqnarray}\label{a3g}
\nonumber
  A_3&=&\frac{\hbar}{eT_{\rm eff}}\int_{-\infty}^{\infty} dt\bigg\{
 - \frac{1}{2}  \left(\frac{\hbar}{2e}\right)^2
\frac{1}{\left(k_BT_{\rm eff}\right)^2}\, \mathcal{C}_3\,
\dot\varphi_2^3
  \\
  \nonumber\\
   &&+\bigg[\frac{\hbar}{e}C\ddot\varphi_2
   -\frac{\hbar}{2e}\frac{1}{R_{\vert\vert}}
   \dot\varphi_2\bigg]\Lambda_3^{\prime}
   \bigg\}
    \, .
\end{eqnarray}

Comparing the form of the evolution equation
(\ref{eqmlamda3strich}) with the one satisfied by $\varphi_2$,
namely Eq.~(\ref{eqmphi}) for $I_3=0$, we are led to the
ansatz\begin{equation} \label{ansatzeom}
\Lambda_3^{\prime}(t)=A(t)\dot\varphi_2(t)\, .
\end{equation}
Inserting this into Eq.~(\ref{eqmlamda3strich}) and using the
evolution equation for $\varphi_2$ as well as Eq.~(\ref{inhom}), 
we find that $A(t)$ obeys the differential equation
\begin{eqnarray}\label{eqma}
\nonumber
  && \frac{\hbar}{2e}C\left[
   2\left(\dot A + \frac{1}{R_{\vert\vert}C}A\right)\ddot\varphi_2
+    \left(\ddot A + \frac{1}{R_{\vert\vert}C}\dot
   A\right)\dot\varphi_2\right]\\
   \nonumber\\
   &&\qquad=-\left(\frac{1}{k_BT_{\rm
   eff}}\right)^2\left(\frac{\hbar}{2e}\right)^2\mathcal{C}_3\,
   \dot\varphi_2\ddot\varphi_2\, ,
\end{eqnarray}
which is satisfied, provided
\begin{equation}\label{eqmaa}
   \dot A + \frac{1}{R_{\vert\vert}C}A=-
   \frac{1}{3C}\frac{\hbar}{2e}
   \frac{1}{\left(k_BT_{\rm eff}\right)^2}\, \mathcal{C}_3\,
   \dot\varphi_2\, .
\end{equation}
When the ansatz (\ref{ansatzeom}) is plugged into (\ref{a3g}), we
obtain a term proportional to $A\dot\varphi_2\ddot\varphi_2$,
which under the integral can be replaced by $-\frac{1}{2}\dot A
\dot\varphi_2^2$. Accordingly, we find
\begin{eqnarray}\label{a3h}
\nonumber
  A_3&=&\frac{\hbar}{eT_{\rm eff}}\int_{-\infty}^{\infty} dt\bigg\{
 - \frac{1}{2}  \left(\frac{\hbar}{2e}\right)^2
\frac{1}{\left(k_BT_{\rm eff}\right)^2}\, \mathcal{C}_3\,
\dot\varphi_2^3
  \\
  \nonumber\\
   &&-\bigg[\frac{\hbar}{2e}C\dot A
   -\frac{\hbar}{2e}\frac{1}{R_{\vert\vert}}A\bigg]
   \dot\varphi_2^2
   \bigg\}
    \, .
\end{eqnarray}
Finally, in the integrand, the expression between squared brackets
can be transformed by means of Eq.~(\ref{eqmaa}) to yield for
$A_3$ the compact result
\begin{equation}\label{a3i}
   A_3= -\frac{k_B}{3}\left(\frac{\hbar}{2e}\right)^3
\frac{1}{\left(k_BT_{\rm eff}\right)^3} \, \mathcal{C}_3
   \int_{-\infty}^{\infty} dt \, \dot\varphi_2^3\, .
\end{equation}

%
%\bibliography{data}

\begin{thebibliography}{10}
\bibitem{remark1}
State-dependent transport coefficients give rise to multiplicative
noise and imply, for instance, differences between deterministic
and average drift. Also the Riemannian curvature of state space
needs to be accounted for. The full complexity of the problem is
treated in the literature\cite{GG,GGG}.
\bibitem{remark2}
The relations from Refs.~\onlinecite{GG,GGG} have been specialized
to a system with state-independent transport coefficients in
contact with a heat bath.
\bibitem{remark3}
In fact, there is a one parameter family of solutions shifted
relative to each other in time. This gives rise to a zero mode in
the spectrum of fluctuations about the minimal action trajectory.
Since in this work we restrict ourselves to the exponential factor
of the rate expression, the fluctuation spectrum will not be
addressed. The solution $\varphi_{\rm relax}(t)$ can be fixed by
requiring in addition, e.g., $\varphi_{\rm
relax}(t)>\frac{\pi}{2}$ for $t<0$ and $\varphi_{\rm
relax}(0)=\frac{\pi}{2}$.
\bibitem{else}
A general account of the theory of irreversible processes and 
non-Gaussian fluctuations in terms of thermodynamically conjugate 
variables will be presented elsewhere.
\bibitem{remark4}
Experimentally, the component $V_N/R_N$ is often eliminated with
the help of an additional branch, see
Refs.~\onlinecite{Pekola,Pothier}. However, since $V_B/R_B$ is an
independently controlled current, the same range of parameters is
accessible.
\bibitem{remark5}
Experimentally, this region can be chosen even at low temperatures
by making the junction capacitance $C$ sufficiently large, in case
of need by means of a parallel shunt capacitor.
\end{thebibliography}
%\bibliographystyle{prsty}

\end{document}